 \documentclass[final,3p,times]{elsarticle}

\usepackage{amssymb}
\usepackage{amsmath}
\newcommand\norm[1]{\lVert#1\rVert} 

\usepackage{bm} 
\usepackage{svg}
\usepackage{svg-extract}
\usepackage{booktabs}
\usepackage{multirow}
\usepackage{setspace} 
\usepackage{lineno}   
\biboptions{sort&compress} 

\journal{Chemical Engineering Journal}

\begin{document}


\begin{frontmatter}



\title{Robust pore-resolved CFD through porous monoliths reconstructed by micro-computed tomography: from digitization to flow prediction}

\affiliation[inst1]{organization={CHAOS, Department of Chemical Engineering},
            addressline={Polytechnique Montreal, PO Box 6079, Stn Centre-Ville}, 
            city={Montreal},
            postcode={H3C 3A7},
            state={Quebec},
            country={Canada}}
\affiliation[inst2]{organization={CREPEC, Department of Chemical Engineering},
           addressline={Polytechnique Montreal, PO Box 6079, Stn Centre-Ville}, 
            city={Montreal},
            postcode={H3C 3A7},
            state={Quebec},
            country={Canada}}            
\affiliation[inst3]{organization={
Department of Chemical and Biotechnological Engineering},
            addressline={Université de Sherbrooke, 2500 Bd de l'Université}, 
            city={Sherbrooke},
            postcode={J1K 2R1},
            state={Quebec},
            country={Canada}}

\author[inst1]{Olivier Guévremont}
\author[inst1]{Lucka Barbeau}
\author[inst2]{Vaiana Moreau}
\author[inst3]{Federico Galli}
\author[inst2]{Nick Virgilio}
\cortext[cor1]{Corresponding author.}
\author[inst1]{Bruno Blais\corref{cor1}}      
\ead{bruno.blais@polymtl.ca}

\begin{abstract}

Porous media are ubiquitous in energy storage and conversion, catalysis, biomechanics, hydrogeology, as well as many other fields. 
These materials possess high surface-to-volume ratios and their complex channels can restrict and guide the flow. However, optimizing design parameters for specific applications remains challenging due to the intricate structure of porous media.
Pore-resolved CFD reveals the effects of their structure on flow characteristics, but is limited by the performance of mesh generation algorithms for such complex geometries. To alleviate this issue, we use a sharp immersed boundary method which enables usage of Cartesian, non-conformal grids, within a massively parallel finite element framework. This method preserves the order convergence of the scheme and allows for adaptive mesh refinement (AMR). We introduce a radial basis function-based representation of solids that allows to solve the flow through complex geometries with precision.
We verify the method using the method of manufactured solutions. 
We validate it using pressure drop measurements through porous silicone monoliths digitized by X-ray computed microtomography, for pore Reynolds numbers up to 30. Simulations are conducted using grids of 200 M cells distributed over 8 k cores, which would require 16 times more cells without AMR. 
Results reveal that pore network structure is the principal factor describing pressure evolution and that preferential channels are dominant at this scale. 
In this work, we demonstrate a robust and efficient workflow for pore-resolved simulations of porous monoliths.
This work bridges the gap between sub-millimetric flow and macroscopic properties, which will open the door to design and optimize processes through the usage of physics-based digital twins of complex porous media.

\end{abstract}


\begin{highlights}
    \item Sharp immersed boundary method using Radial Basis Functions (RBF) in parallel simulations ($>8$k cores).
    \item Fully pore-resolved flow through porous monoliths (pore size: 300 $\mu$m, volume: 0.3 mL).
    \item CFD through any mesh-defined solid using non-conformal grids in a finite element framework.
    \item Efficient workflow from monolith digitization to resolved velocity and pressure fields.
    \item Robust flow verification around arbitrary shapes using the method of manufactured solutions.
\end{highlights}

\begin{keyword}
Pore-resolved CFD \sep
Radial basis functions \sep
Immersed boundary \sep
Porous media \sep
Finite element method \sep
High-performance computing
\end{keyword}

\end{frontmatter}


\section{Flow in porous media based on microtomography}
\label{sec:littrev}

Porous media are attractive materials for a wide variety of application fields: heterogeneous catalysis \cite{gavriilidis2002microreactorsApplications,zhang2019emergingPorousChromato}, separation (chromatography) \cite{zhang2019emergingPorousChromato}, biomedical engineering (interstitial flows, drug delivery) \cite{swartz2007interstitial,khaled2003bioPorous}, filtration \cite{yadroitsev2009PorousFiltration}, and hydrogeology \cite{dippenaar2014hydrogeology_porosity}, to name a few. 
They usually exhibit high specific surface and  channel networks that can form tortuous, merging and forking paths that limit the thickness of boundary layers, depending on properties such as porosity, pore size distributions, tortuosity and isotropy. 
Continuous flow micro-reactors, in which channels are sub-millimetric, take advantage of the low transport distances resulting from narrow channels, as well as a fine control of the process parameters \cite{dong2021defineMicroReactor}. 
Along with the commonly studied residence time distribution and pressure drop, accurately predicting the velocity profile in these sub-millimeter channels is essential for the design of these equipment.
The mixing effects and boundary layers must be resolved in order to predict and control mixing, chemical reactions, and other transport phenomena.
However, obtaining quality data regarding the velocity field is challenging. Experimental methods such as Magnetic Resonance Velocimetry (MRV) \cite{ricke2023magneticVelocimetry} or Particle Image Velocimetry (PIV) \cite{lu2018PIVmicroTransparent} are limited: MRV is limited by spatial resolution and PIV by the opacity of the medium.

Given these challenges, Computational Fluid Dynamics (CFD) has been used for decades to solve the pore-resolved flow in porous media \cite{golparvar2018review_multiphase_porous,wood2020review_turbFlo_porousmedia} (PRCFD). At first, most of the work published on the subject considered the flow in idealized and periodic Representative Elementary Volumes (REV) to reduce computational costs \cite{saxena2018rev_considerations_imaging}, but this approach is limited to scale-independent domains, simple geometries and is complex to validate experimentally. Lately however, increases in hardware and software capabilities gave rise to more investigations coupled with realistic media considering all of their heterogeneities. Digitization techniques such as focused ion beam scanning electron microscopy (FIB-SEM) \cite{curtis2010fibsem}, X-ray microcomputed tomography ($\mu$CT) \cite{flannery1987xct} and nuclear magnetic resonance (NMR) \cite{chen2022porescalemodelling_review,rychlik2004acquisition} have been used to obtain stacks of images representing complex media in a voxelized manner. These
voxelized representations 
have both been used as-is with the Lattice Boltzmann method (LBM) \cite{liu2015lbm_review,sanematsu2015xct_lbm_fem}, or have been used to construct volume meshes of the phases, to be simulated with the Finite Volume method (FVM) \cite{ranut2014opencellCFD_fvmansys,piller2009ct_FVM_DNS,aboukhedr2017ctcfd_vof_droplet,emmel2020ctcfd_starccm_electrolyte,sinn2020cht_ctcfd,dong2018ctcfd_fixedbedreactor_ht_species} or Finite Element method (FEM) \cite{menegazzi1997enginecoolingCT_CFD_FEM,roberts2016tetraedralFEM,sanematsu2015xct_lbm_fem}.

Starting from voxelized representations of real porous media has the advantage of keeping a realistic morphology (surface, volume, structure), but requires multiple pre-treatment steps when using FVM or FEM including \cite{kuhlmann2022_microCT_to_CFD}: (1) digitization, (2) segmentation of phases, (3) creation of a surface mesh, (4) decimation and smoothing of the surface mesh, and (5) generation of volume meshes.
While recent work involving the digitization of porous media to run CFD simulations have used most of these steps, Kuhlmann et al. \cite{kuhlmann2022_microCT_to_CFD} highlighted the complexity of the process, the fact that strings of software need to be used to produce a mesh, and how expertise both in digital image processing and CFD is a rare occurrence. This inhibits usage and advancement in the field of PRCFD.

An alternative path to account for the complex morphologies of porous media in CFD is to avoid conformal meshes altogether, and instead use immersed boundary methods (IBM) \cite{das2018multiscaleOpencellfoam,das2018sharpIB,das2016bImmersedBoundary,chandra2019opencellfoams}. 
This method of representing a solid embedded in a fluid was first introduced by Peskin \cite{peskin1972IB} and was based on the addition of forcing terms in the Navier-Stokes equations.
IBM have been implemented and used with packings of particles both for Finite Difference method (FDM) \cite{lu2018direct,lu2019direct} and FVM \cite{chandra2020dns,chandra2021fishertropsch}. Das et al. \cite{das2018multiscaleOpencellfoam} have used IBM to solve the flow through packings of non spherical particles. To our knowledge, IBM have been used for flow in porous media obtained from digitized media only by Das et al. \cite{das2016bImmersedBoundary} (FVM),  Ilinca and Hétu \cite{ilinca2010cutFEM,ilinca2011cutFEMstatic,ilinca2012IB_fem_porous} (FEM) and Lesueur et al. \cite{lesueur2022unfittedFEM_CT} (FEM).
Das et al. \cite{das2016bImmersedBoundary} used a sharp IBM to accurately account for the solid represented by a surface mesh when solving the flow.
Ilinca and Hétu \cite{ilinca2010cutFEM,ilinca2011cutFEMstatic,ilinca2012IB_fem_porous} solved the flow using the cut cell method by adding degrees of freedom on the embedded surface, creating a conformal mesh locally during assembly before these degrees of freedom are removed by static condensation.
Lesueur et al. \cite{lesueur2022unfittedFEM_CT} moved and fitted the existing nodes of cut cells from a non-conformal grid to the solid surface before assembling the system based on this newly formed conformal mesh.

We believe that the scarcity of works starting from $\mu$CT to simulate the flow in porous media is also due to the prohibitive 
cost of evaluating the distance to highly detailed objects obtained from digitization.
When solving the flow using IBM, the distance between each discretization point and the solid surface must be computed repeatedly, which is not viable when the solid is represented as a collection of millions of voxels or polygons, as is the case for porous media.
To alleviate this issue, we propose a Radial Basis Function (RBF) network approach in which an approximate signed distance function (SDF) of the solid is encoded, as described by Liu et al. \cite{liu2019parallelRBF}. RBF networks are suitable to represent complex functions in space \cite{poggio1990rbfUnivApprox,park1991rbfUnivApprox,liao2003rbfUnivApprox} and have been used to approximate distance fields accurately \cite{carr2001reconstruction}. In addition, they provide complete support of the approximated functions in the vicinity of the nodes and can be used with compact basis functions, rendering RBF-encoded shapes portable and efficient in massively parallel simulations \cite{liu2019parallelRBF,biancolini2017fast}.
RBF networks representing solids have been used for structural topology optimization \cite{liu2019parallelRBF} and CFD around simple shapes, including: cylinders \cite{toja2024cfd_ib_rbf,thai2013compactRBF2d_direct}, a NACA0012 airfoil \cite{xin2018rbf_ib}, and platelets (blood) \cite{shankar2015augmentingRBF}. However, the application of RBF to represent complex porous media and its coupling to large scale CFD simulations has not been demonstrated.

The purpose of this work is to extend the applicability of IBM to complex static solids (e.g. porous media) in a flexible, parallelized and accurate manner. 
We introduce a novel method of
integrating 
RBF networks in a CFD framework that is based on distributed grids. We verify the implementation and workflow using a type of manufactured solutions that are applicable to any geometry. Finally, we formulate and validate experimentally a workflow to predict the flow field across real digitized porous monoliths in their entirety, focusing on flows at pore Reynolds number values below 30. Applications of such flows include environmental engineering (groundwater flow models \cite{medici2021reviewGroundwater,wang2019experimentalGroundwater}), filtration (e.g. membrane filtration \cite{wang2010membraneFiltration}), heat exchange \cite{mehrizi2013porousHeatExchangelowRe}, micro-reactors \cite{dong2021defineMicroReactor}, and fuel cells \cite{barreras2005fuelCelllowRe}. 
The result is a physics-based digital twin of a real porous medium that can predict both sub-millimetric flow characteristics from numerical solid representations, and provide macroscopic averaged properties that are experimentally measurable.

\section{Methodology}
\label{sec:methodology}
\begin{figure}[htpb!]
        \centering
        \includegraphics[width=\textwidth]{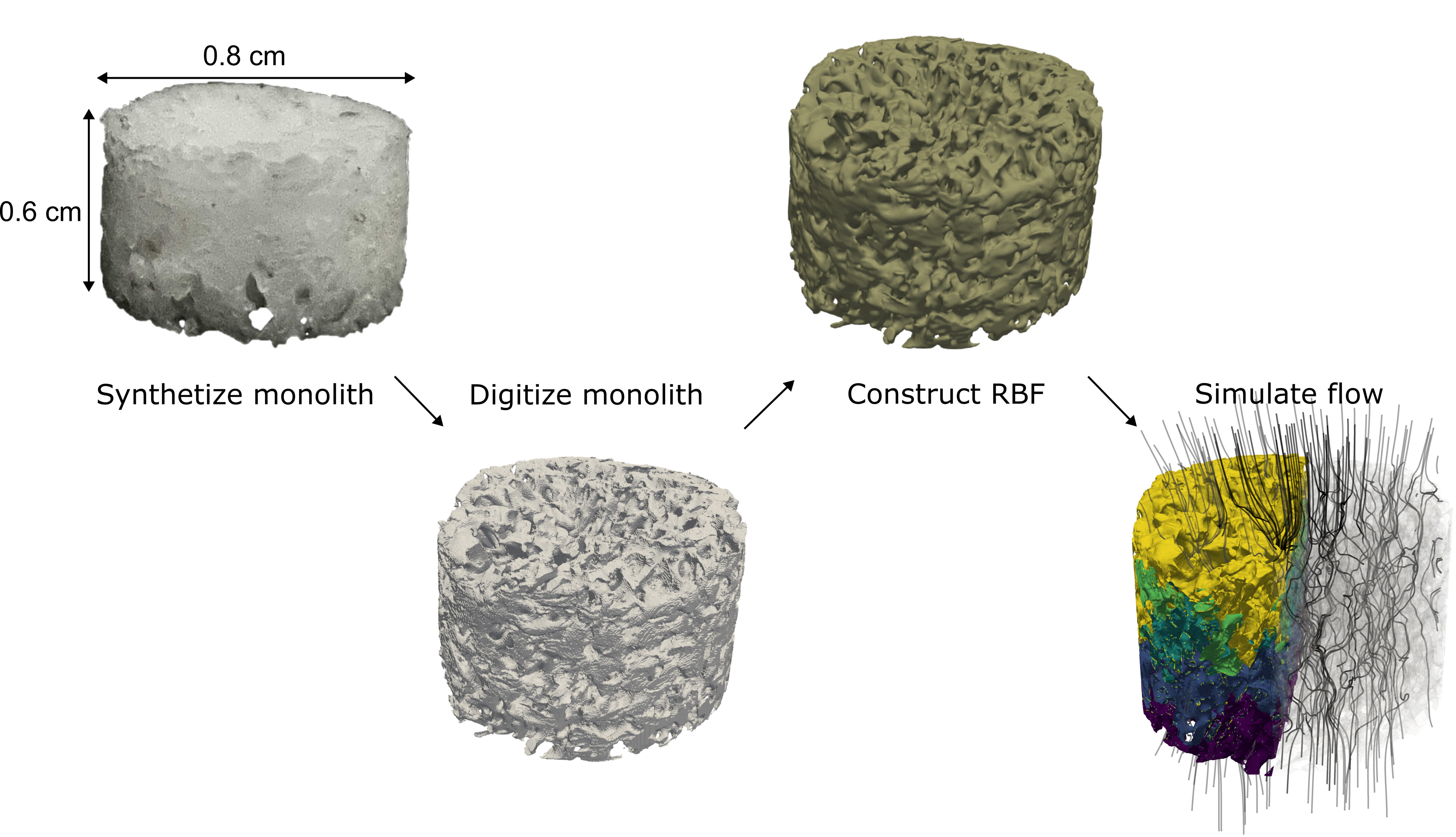}
        \caption{Process of synthetizing a monolith (\ref{sec:porous_samples_synthesis}), digitizing this monolith (Section \ref{sec:section_tomo}), constructing a RBF-network representation of its shape (Section \ref{sec:rbf_network_training}) and simulating the flow through its pores (Section \ref{sec:sim_setup}). The flow simulation results show the monolith colored by the normalized pressure field at the surface, with the streamlines following the velocity field.}
        \label{fig:digitization_workflow}
    \end{figure}

To simulate the flow through real porous monoliths, we synthetize silicone monoliths (\textbf{\ref{sec:porous_samples_synthesis}}), digitize them (\textbf{Section \ref{sec:section_tomo}}), and construct RBF-network approximations of their geometry (\textbf{Section \ref{sec:rbf_network_training}}). We verify the method and its orders of convergence (\textbf{Section \ref{sec:verification}}). We measure experimentally the pressure drop at various flow rates (\textbf{Section \ref{sec:validation_exp}}) to compare to values obtained by simulation (\textbf{Section \ref{sec:sim_setup}}), 
for validation. The workflow (excluding verification and validation) is shown in \textbf{Figure \ref{fig:digitization_workflow}}.

\subsection{Finite element formulation and immersed boundary method \label{sec:fem}}
    We use the FEM to solve the flow of the fluid phase by solving the incompressible Navier-Stokes equations, where $\bm{u}$ is the velocity, $p$ is the pressure, $\rho$ is the density, $\bm{\tau}$ is the deviatoric stress tensor, $f$ is a source term and $\nu$ is the kinematic viscosity. We include the density in the pressure definition and use $p^* = \frac{p}{\rho}$.
    \begin{flalign}
        \nabla \cdot \bm{u} &= 0 \\
        \frac{\partial \bm{u}}{\partial t} + (\bm{u} \cdot \nabla) \bm{u} &= - \nabla p^* + \nabla \cdot \bm{\tau} + \bm{f}
    \end{flalign}
    \begin{equation}
        \bm{\tau} = \nu \left( (\nabla \bm{u}) + (\nabla \bm{u})^T \right)
    \end{equation}
    
    The solver used in this work is part of the open-source software Lethe \cite{blais2020lethe}, based on the deal.II library \cite{bangerth2007dealii}. It uses distributed adaptive meshes from p4est \cite{p4est2011} to enable simulations over a large number of cores ($>10$k). Lethe uses a streamline upwind Petrov–Galerkin and pressure-stabilizing Petrov-Galerkin (SUPG/PSPG) stabilized finite element formulation to solve the Navier-Stokes equations, which allows to use the same order of elements for velocity and pressure \cite{blais2020lethe}. We use linear cubic elements for both velocity and pressure (Q$_1$Q$_1$), as well as quadratic-linear elements (Q$_2$Q$_1$) for verification. Transient simulations are realized using a backward difference formula (BDF) of first order \cite{hay2015bdf_timeintegration}.
    
    We use more specifically Lethe's Sharp-interface Immersed Boundary Method (SIBM), which was used to simulate the flow around spherical particles \cite{barbeau2022}, the flow-particle coupling of multiple particles \cite{barbeau2024,barbeau2024rom} and the non-Newtonian flow around a sphere \cite{daunais2023}. We use the SIBM to account for complex geometries while avoiding conformal mesh generation. SIBM requires a SDF and its gradient to properly apply the constraints to the discretized Navier-Stokes equations. The steps to obtain the SDF of a porous medium are described in the next subsections.

\subsection{Digitization\label{sec:digitization}}
    We synthetize porous silicone samples (cylinders of  $6$ mm height and $8$ mm diameter) following the method described in \ref{sec:porous_samples_synthesis}, then digitize them using $\mu$CT.
    Even though the same protocol was used for each sample, they each have a different morphology due to the stochasticity of polymer blend extrusion and the subsequent coarsening process.
    We segment the data to generate a surface mesh representing the solid, then train RBF networks to approximate the SDF around the digitized objects. The RBF-encoded SDF can then be used to represent 
    the surface of the porous media implicitly at the zeroth isovalue.
    The $\mu$CT and RBF-network training steps correspond to the second and third steps in Figure \ref{fig:digitization_workflow}.
    
    \subsubsection{X-ray microcomputed tomography \label{sec:section_tomo}}
    We digitize all samples using $\mu$CT analysis with a ZEISS Xradia 520 Versa XRM instrument. We position each sample in the support used for the pressure drop experiments before the scan to ensure that the potential compression of pores is considered within the $\mu$CT scan. We set the voxel size of the captured images to be at most $10$ $\mu$m. We use Dragonfly Pro Version 2020.1, Build 809 (Object Research Systems, Inc, Montreal, Quebec, Canada) to segment the phases (solid and pores). We apply a Gaussian filtering to the image, then identify the solid region visually to define a Region Of Interest (ROI). We set the complementary region as the pores. 
    We generate a smoothed surface mesh on the solid ROI after selecting a threshold parameter value, which removes the staircase-like effect associated to the voxelized representation of digitized media.
    As highlighted in the biomedical literature \cite{stock2020pelvicCT, badriyah2019brainCT,friedli2020craniumCT}, the segmentation procedure can have a significant effect on the results obtained from subsequent analyses, especially for threshold-based segmentation. The threshold parameter is related to the intensity of voxel coloration, and must be set between 0 and 100 \% in Dragonfly. As the selected value increases, the resulting representation of the solid goes from full to more skeletal. Care must be exerted since the level of porosity can have a significant effect on the pressure drop. Given the lack of information for our specific combination of materials, $\mu$CT equipment and settings, we use threshold values of $\{$1, 50, 99$\}$ \% when generating surface meshes from solid ROIs to establish
    the sensitivity on the threshold
    of the solid reconstruction and, consequently, of the pressure drop.

    \subsubsection{Characterization of silicone monoliths}
        We use Dragonfly and its integrated workflow for pore network analysis using OpenPNM \cite{gostick2016openpnm}. OpenPNM generates networks by using PoreSpy's Sub-Network of an Over-segmented Watershed algorithm \cite{gostick2019porespy,gostick2017openpnm_watershed}.
        The pore size and distance between connected pores distributions obtained are shown in \textbf{\ref{sec:appendix_pore_size_distributions}}.
        The mean pore size (by volume) and standard deviation reported in \textbf{Table \ref{tab:samples_characterization}} are based on the distribution of inscribed diameter (largest sphere that can fit in a pore), under the assumption of normal pore size distribution. The distance between pairs of pores is computed between their centroids under the same asssumption.

        We compute the specific surface and porosity using the voxel volume and surface from the solid and pores ROI, and correct the surface by a factor $2/3$, as proposed by Yeong and Torquato \cite{yeong1998a,yeong1998b}, to account for the surface voxelization into a staircase-like fashion of digitized random porous structures.
        
        \begin{table}[htpb!]
            \centering
        	\caption{Porous characteristics of each sample: mean pore size $d_p$ (by volume), pore size standard deviation $\sigma_p$ (by volume), mean distance between connected pores $d_c$ (by number), distance between connected pores standard deviation $\sigma_c$ (by number), specific surface $S_p$ and porosity $\varepsilon$.
         }
        	\label{tab:samples_characterization}
        	\begin{tabular}{c c c c c c c}
        		\hline
        		\textbf{\#} &  $d_p$ $ [\mu m]$ & $\sigma_p$ $[\mu m]$  &  $d_c$ $ [\mu m]$ & $\sigma_c$ $[\mu m]$ & $S_p$ $[\textit{cm}^{-1}]$ & $\varepsilon$ $[\%]$  \\
        		\hline
        		\hline
        		  1 & 276 & 93 & 417 & 175 & 68 & 53 \\ 
        		  2 & 346 & 80 & 440 & 189 & 50 & 50 \\ 
        		  3 & 366 & 132 & 530 & 199 & 60 & 55 \\ 
        		\hline
        	\end{tabular}
        \end{table}
    
    \subsubsection{RBF network training}    
        \label{sec:rbf_network_training}
        We use a Radial Basis Function (RBF) network generation tool based on the open-source Bitpit library \cite{bitpit}, which takes advantage of its Parallel Balanced Linear Octree (PABLO), Levelset and RBF capabilities. 
        The RBF network ($\tilde{\lambda}$) approximates the SDF ($\lambda$) of the shape by evaluating the contribution of each RBF-node based on the evaluation point ($\bm{x}$), basis functions ($\varphi_i$) and weights ($\gamma_i$), obtained from training. The expression of the RBF network is:
    
        \begin{equation}
            \lambda(\bm{x}) \approx \tilde{\lambda}(\bm{x}) = \sum_i \varphi_i (\bm{x}) \gamma_i
            \label{eq:rbf_network}
        \end{equation}
        
\begin{figure}[htpb!]
        	\centering
        	\includegraphics[width=\textwidth]{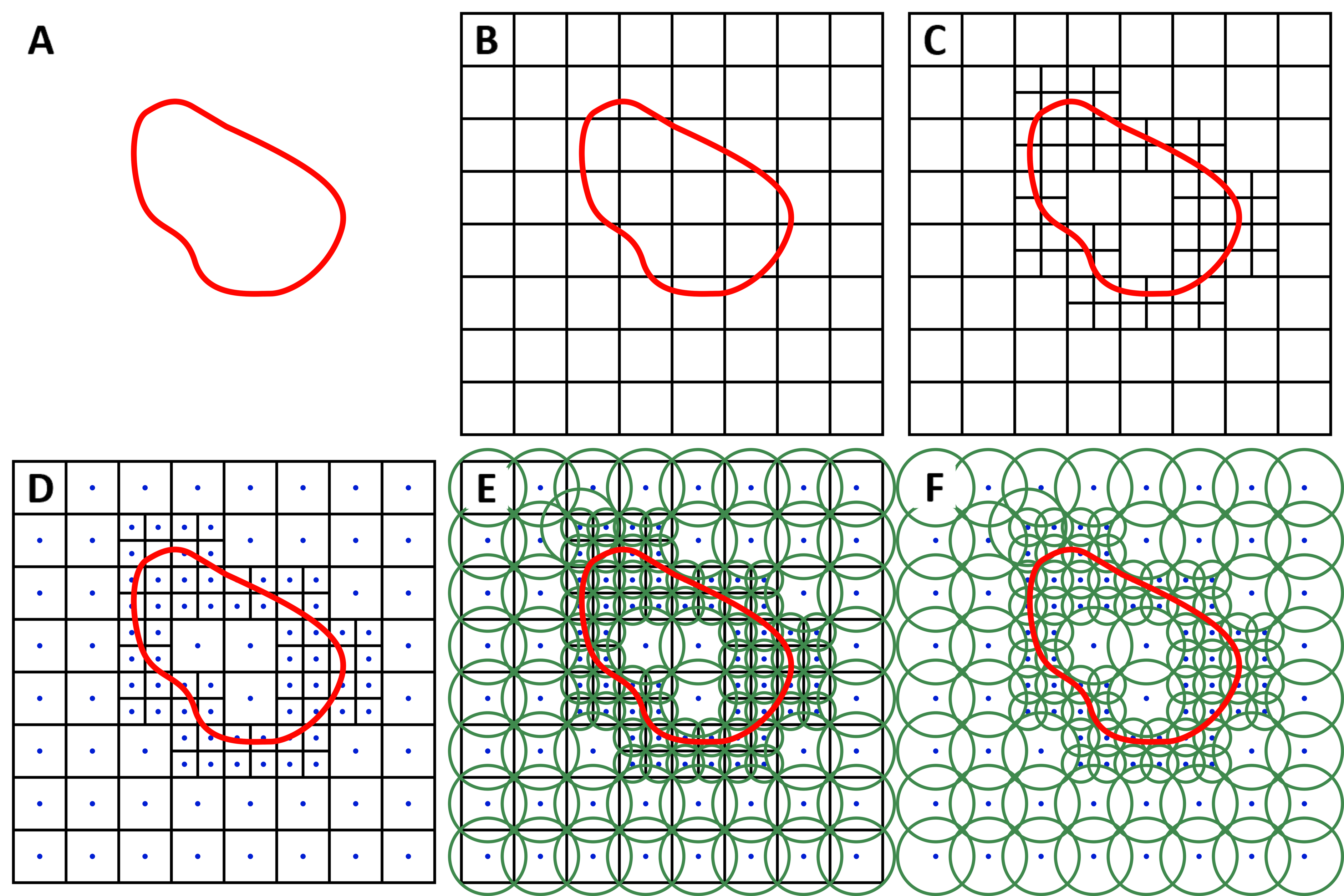}
        	\caption{Steps used to place RBF-nodes around an object and associate compact basis functions to each node. (\textbf{A}) Around a given object, (\textbf{B}) a uniform grid is formed (\textbf{C}) then adaptively refined near the boundary. (\textbf{D}) Nodes are placed at the center of each cell, (\textbf{E}) then a basis function is associated to each node. (\textbf{F}) The grid is not needed anymore.}
        	\label{fig:potatoid_rbf}
        \end{figure}

        The training steps are:
        \begin{enumerate}
            \item Load the STereoLithography (STL) file representing the porous medium obtained from $\mu$CT (section \ref{sec:section_tomo});
            \item Build a uniform cartesian grid of $(2^m)^3$ cells around it, based on a bounding box $p$ percent larger than the object. 
            We will generate RBF-nodes at each cell centroid; these will also be used as collocation points. This step and subsequent ones are illustrated in \textbf{Figure \ref{fig:potatoid_rbf}}.
            \item (Optional) Refine adaptively the cells cut by the object. Repeat this step $n$ times to obtain the desired level of detail;
            \item Assemble a system of equations structured as $\bm{A} \bm{\gamma} = \bm{b}$. 
            The elements of $\bm{A}$ and $\bm{b}$ are respectively $a_{i,j} = \varphi_j (\bm{x}_i)$ and $b_i = \lambda(\bm{x}_i)$, computed from the surface mesh. 
            Each row $i$, corresponding to a cell centroid, is built as shown in \textbf{Equation \ref{eq:rbf_training}}, where $j$ indicates the column of the matrix.
            \begin{equation}
                \lambda(\bm{x}_i)  = \sum_j \varphi_j (\bm{x}_i) \gamma_j
                \label{eq:rbf_training}
            \end{equation}
            The matrix $\bm{A}$ is sparse when using compact support for the basis functions;
            \item Solve the system to obtain the weights $\gamma_i$ using a Least Squares Conjugate Gradient (LSCG) solver, because $A$ is poorly conditioned when adaptive refinement is used. Conjugate Gradient (CG) solvers can be used otherwise. 
            \item Export the resulting RBF network.
        \end{enumerate}

        The number of RBF-nodes can be minimized using
        a higher density of nodes close to the zeroth isovalue of the SDF. This ensures that the main features of the shapes are captured, while tolerating a higher error on the SDF further from the zeroth isovalue. 
            
        We choose basis functions 
        which have a compact support to ensure a sparse linear system of equations.
        We use the C2 Wendland basis function for its distance approximation capabilities \cite{wendland1995piecewise} and because it is twice continuously differentiable: this property is desirable for the SIBM which requires the gradient of the SDF \cite{barbeau2022,barbeau2024}. 

        \begin{equation}
            \varphi_\text{i,Wendland} (\bm{x}) =  
            \begin{cases}
                \left(1 - \frac{\norm{\bm{x} - \bm{x_i}}}{\kappa_i}\right)^4 \left(4 \frac{\norm{\bm{x} - \bm{x_i}}}{\kappa_i} + 1\right) ,& \text{if } \frac{\norm{\bm{x} - \bm{x_i}}}{\kappa_i}\leq 1\\
                0,              & \text{otherwise}
            \end{cases}
            \label{eq:wendland}
        \end{equation}

        The support radius of each RBF-node is $\kappa_i = \alpha h_i $, where $\alpha$ is a user-defined constant and $h_i$ is the cell length. 
        This definition of $\kappa_i$ provides a way to use compact basis functions while capturing the features of the media. 
        We use $\alpha = 3$ for this work since it is a good compromise between speed, efficiency and accuracy.
        Training is carried out on the Digital Research Alliance of Canada's cluster Narval, on a single node using 2 sockets with 32 AMD Rome 7532 (2.40 GHz 256M cache L3) processors and 249 GB of Random Access Memory.

\subsection{RBF networks for immersed boundary application\label{sec:rbf}}
    
    \subsubsection{Initialization}
        In Lethe, the initialization ensures that each cell has access to a list of the RBF-nodes whose contribution is possibly non-null over the cell, allowing us to remove superfluous data and speedup the RBF network usage. 
        This step takes advantage of the parallel distributed hierarchical structure of the meshes within Lethe. 
        From the coarsest level to the finest, RBF-nodes that do not contribute to the RBF network in locally owned cells are discarded. Starting from the coarsest level and moving information upwards allows to filter relevant RBF-nodes efficiently for each parallel process.
        
    \subsubsection{Evaluation}
        We only sum over RBF-nodes whose support radius overlaps with the evaluation point. The SDF and its gradient are: 
        \begin{equation}
            \lambda(\bm{x}) \approx 
            \sum_{i}  \varphi_i (\bm{x}) \gamma_i
        \end{equation}

        \begin{equation}
            \nabla \lambda(\bm{x}) \approx  \sum_{i}  \nabla \varphi_i (\bm{x}) \gamma_i
        \end{equation}
        
        To reduce the computational cost, a caching mechanism is used to rapidly sum over the RBF network.

\section{Verification}
\label{sec:verification}
The SIBM implemention in Lethe has been extensively verified by Barbeau et al. \cite{barbeau2022,barbeau2024, barbeau2024rom} and Daunais et al. \cite{daunais2023}. It preserves the order of convergence associated to the underlying FEM scheme. 
In this section, we introduce and use an approach to manufacture solutions around arbitrary shapes.
We focus specifically on the verification of the generalization of the SIBM to 
superquadrics (with implicit \textbf{Equation \ref{eq:superquadric}}) and the same geometries, defined by RBF networks. 
In Equation \ref{eq:superquadric},
\{$a$, $b$, $c$\} are the half-lengths and \{$p$, $q$, $r$\} are the blockiness factors. 
$d$ represents the levels of the implicit superquadric function, with the zeroth isovalue corresponding to the surface itself where $d(\bm{x}_\text{surface})=\lambda(\bm{x}_\text{surface})=0$.
It must be noted that while $d(\bm{x}_\text{surface}) = 0$, $d$ is not a SDF function \textit{per se} as $\left |\nabla d(\bm{x}) \right | \neq 1$. 

\begin{equation}
    d(\mathbf{x}) =  \left | \frac{x}{a} \right |^p + \left |\frac{y}{b} \right|^q + \left |\frac{z}{c} \right |^r  - 1 
    \label{eq:superquadric}
\end{equation}

\subsection{Method of manufactured solutions}
We use the Method of Manufactured Solutions (MMS) \cite{oberkampf2010vv} to verify the proper implementation of the SIBM to analytically defined superquadrics and their RBF network approximations. The SIBM in Lethe uses the SDF to locate the boundary and subsequently impose the proper constraints to the system. We use a similar idea to construct the manufactured solutions for velocity: we multiply base velocity functions by a function that is zero at the surface (e.g. $d$, $\lambda$) to obtain a manufactured solution that has zero velocity at the solid surface. This approach can manufacture solutions for any SDF.

The base solutions are, for $u_b$, $v_b$, $w_b$ and $p$ being the three components of velocity and the pressure, respectively:
\begin{flalign}
    u_b &= \sin^2(\pi x) \cos(\pi y) \sin(\pi y) \cos(\pi z) \sin(\pi z) \\
    v_b &= \cos(\pi x) \sin(\pi x) \sin^2(\pi y) \cos(\pi z) \sin(\pi z) \\
    w_b &= -2 \cos(\pi x) \sin(\pi x) \cos(\pi y) \sin(\pi y) \sin^2(\pi z) \\
    p &= \sin(\pi x) + \sin(\pi y) + \sin(\pi z)
\end{flalign}

The manufactured solution for a superquadric shape is:
\begin{flalign}
    u &= \left ( \left | \frac{x}{a} \right |^p + \left |\frac{y}{b} \right|^q + \left |\frac{z}{c} \right |^r  - 1 \right) 
    \sin^2(\pi x) \cos(\pi y) \sin(\pi y) \cos(\pi z) \sin(\pi z) \\
    v &= \left ( \left | \frac{x}{a} \right |^p + \left |\frac{y}{b} \right|^q + \left |\frac{z}{c} \right |^r  - 1 \right) 
    \cos(\pi x) \sin(\pi x) \sin^2(\pi y) \cos(\pi z) \sin(\pi z) \\
    w &= \left ( \left | \frac{x}{a} \right |^p + \left |\frac{y}{b} \right|^q + \left |\frac{z}{c} \right |^r  - 1 \right) 
    \left (
    -2 \cos(\pi x) \sin(\pi x) \cos(\pi y) \sin(\pi y) \sin^2(\pi z) 
    \right )
    \\
    p &= \sin(\pi x) + \sin(\pi y) + \sin(\pi z)
\end{flalign}
It must be noted that the velocity field is not solenoidal, thus a source term must be added to the continuity equation.

\begin{figure}[htpb!]
	\centering
	\includegraphics[width=\textwidth]{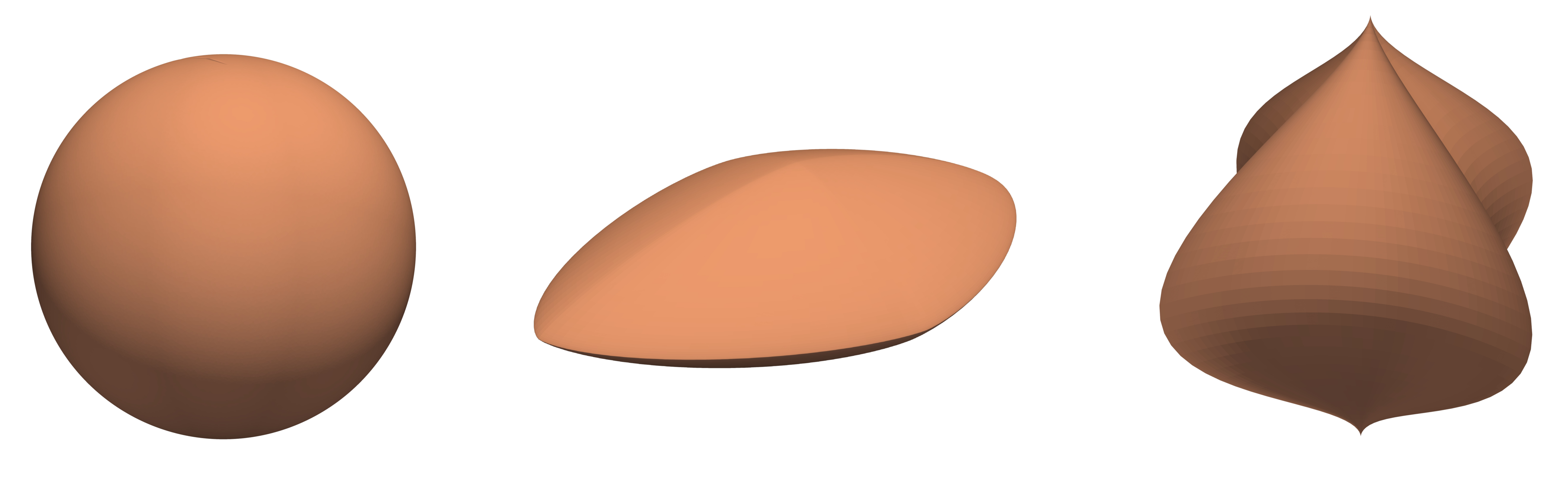}
	\caption{Shapes used for the verification of the flow around superquadric and RBF-encoded shapes.}
	\label{fig:stl_png_all_shapes}
\end{figure}

Three shapes are considered for verification: a sphere, a convex superquadric, and a concave superquadric. Their appearances and associated parameters are shown in \textbf{Figure \ref{fig:stl_png_all_shapes}} and \textbf{Table \ref{tab:parameters_all_shapes}}. 

\begin{table}[htpb!]
	\centering
	\caption{Parameters used to generate the shapes for verification.}
	\label{tab:parameters_all_shapes}
	\begin{tabular}{l c c}
		\hline
		\textbf{Shape} & \textbf{Parameter} & \textbf{Value} \\
		\hline
		\hline
		  Sphere & Radius & 0.50 \\ 
		\hline
		  Convex superquadric & a & 0.50 \\ 
		                      & b & 0.75 \\ 
		                      & c & 0.25 \\ 
		                      & p & 1.50 \\ 
		                      & q & 1.50 \\ 
		                      & r & 1.20 \\ 
		\hline
		  Concave superquadric& a & 0.50 \\ 
		                      & b & 0.50 \\ 
		                      & c & 0.50 \\ 
		                      & p & 0.60 \\ 
		                      & q & 0.60 \\ 
		                      & r & 2.00 \\ 
		\hline
	\end{tabular}
\end{table}

\subsection{Superquadrics and uniform RBF}

\begin{figure}[htpb!]
	\centering
	\includegraphics[width=0.9\textwidth]{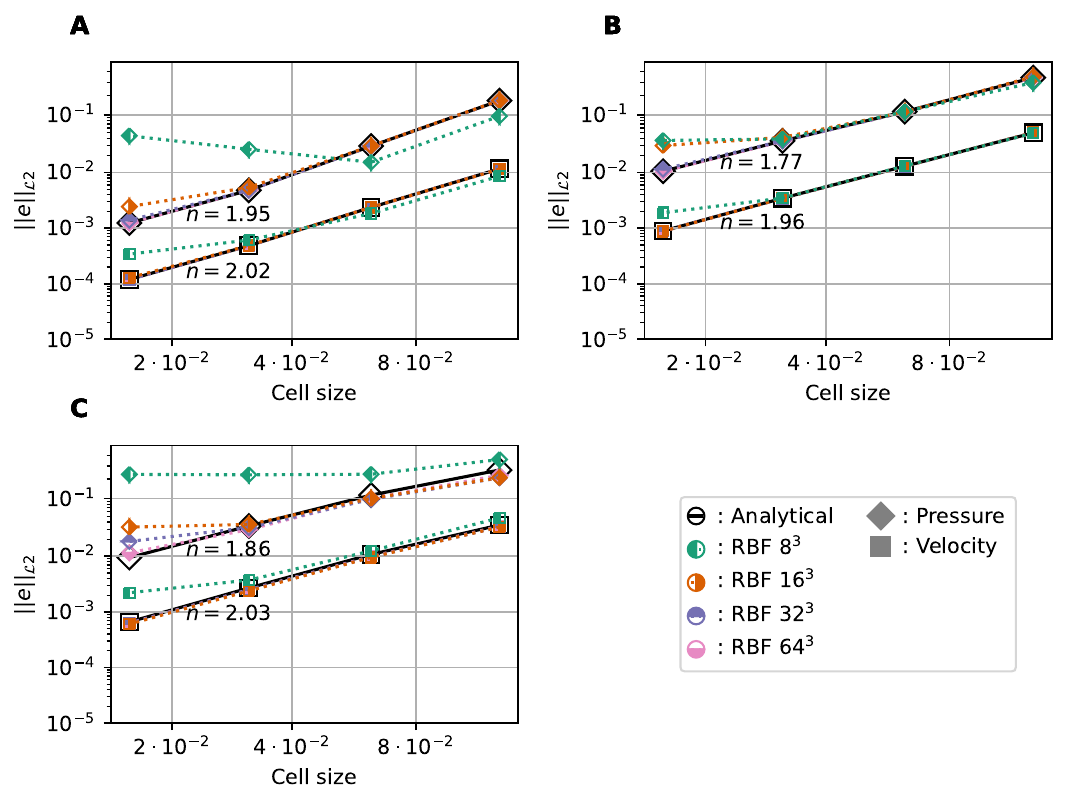}
	\caption{$\mathcal{L}2$ norm of the error of the MMS for the velocity and pressure fields  with Q$_1$Q$_1$ elements around \textbf{A} a sphere, \textbf{B} a convex superquadric (Conv.Sup.) and \textbf{C} a concave superquadric (Conv.Sup.), both for analytical (full lines) and for RBF representations (dotted lines). Diamond and square markers signify pressure and velocity errors, respectively. For RBF representations, markers are half-filled in varying orientations depending of the number of RBF-nodes. $n$ is the order of convergence for the last two grids.}
	\label{fig:mms_combined_Q1Q1}
\end{figure}

\begin{figure}[htpb!]
	\centering
	\includegraphics[width=0.9\textwidth]{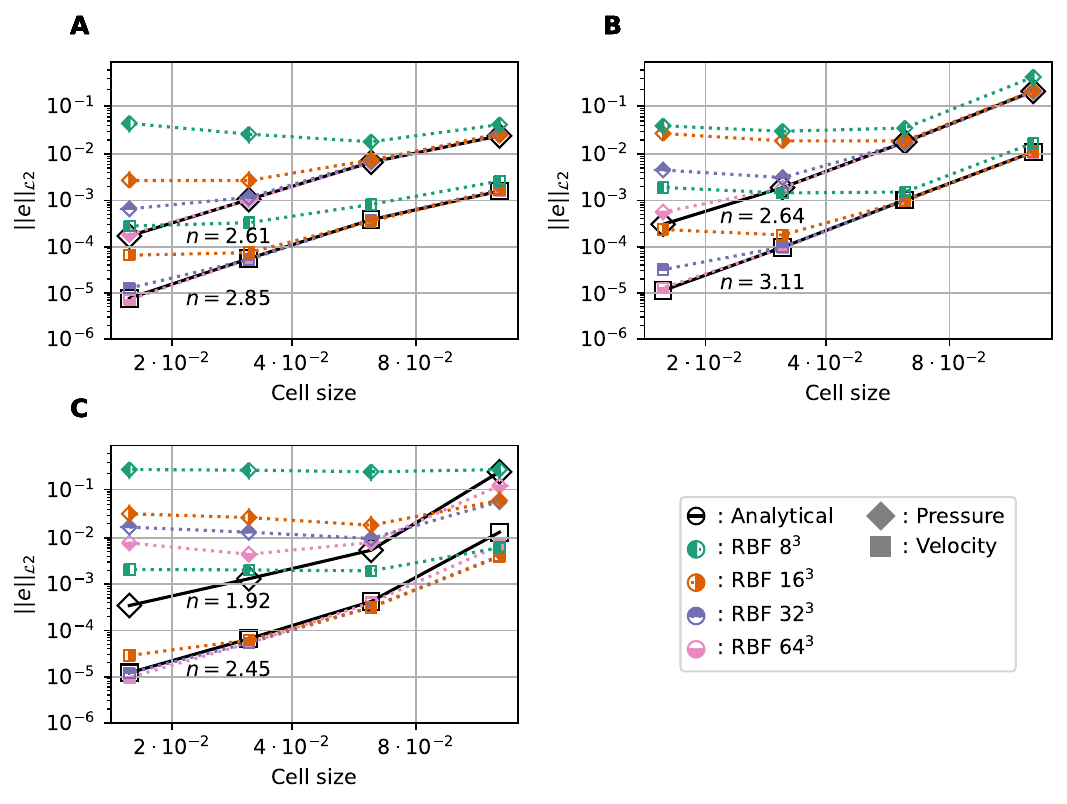}
	\caption{$\mathcal{L}2$ norm of the error of the MMS for the velocity and pressure fields  with Q$_2$Q$_1$ elements around \textbf{A} a sphere, \textbf{B} a convex superquadric (Conv.Sup.) and \textbf{C} a concave superquadric (Conv.Sup.), both for analytical (full lines) and for RBF representations (dotted lines). Diamond and square markers signify pressure and velocity errors, respectively. For RBF representations, markers are half-filled in varying orientations depending of the number of RBF-nodes. $n$ is the order of convergence for the last two grids.}
	\label{fig:mms_combined_Q2Q1}
\end{figure}

We verify the SIBM in two steps, with their results shown in \textbf{Figures \ref{fig:mms_combined_Q1Q1}} and \textbf{\ref{fig:mms_combined_Q2Q1}}. Figure \ref{fig:mms_combined_Q1Q1} shows the decrease of the L2-norm of the error of velocity ($\norm{e_u}_{\mathcal{L}2}$) and pressure ($\norm{e_p}_{\mathcal{L}2}$) in a simulated domain ($[0,1 ]^3$) when using Q$_1$Q$_1$ elements. 
Figure \ref{fig:mms_combined_Q2Q1} shows the same information when using Q$_2$Q$_1$ elements. In this section, RBF representations are built without adaptive refinement (see \ref{sec:rbf_network_training}).

\begin{equation}
    \norm{e_u}_{\mathcal{L}2} = 
    \left (
        \int_\Omega 
            \norm{\bm{u}_\text{analytical} - \bm{u}_\text{approximated}}^2 
            d \Omega
    \right )^{\frac{1}{2}}
    \label{eq:l2_velocity}
\end{equation}

\begin{equation}
    \norm{e_p}_{\mathcal{L}2} = 
    \left (
        \int_\Omega 
            \norm{p_\text{analytical} - p_\text{approximated}}^2 
            d \Omega
    \right )^{\frac{1}{2}}
    \label{eq:l2_pressure}
\end{equation}

\textit{First}, we verify the implementation of the superquadrics class of shapes. The velocity and pressure L2-errors for the flow around superquadrics are represented by full lines. We calculate the orders of convergence $n$ for the last two grids. We notice that the asymptotic convergence region is reached for 
all superquadrics when Q$_1$Q$_1$ elements are used. In the case of Q$_2$Q$_1$ elements, concave superquadrics lead to a slightly inferior convergence rate. This is most likely due to the presence of the singularities and/or sharp features in the geometry.

\textit{Second}, we verify that RBF-encoded shapes that represent superquadrics (dotted lines) recover the same orders of convergence obtained using analytical superquadrics. By increasing the number of RBF nodes used from $8^3$ to $64^3$, the results for both velocity and pressure errors collapse gradually on the analytical superquadrics curves. It is quite apparent, when using similar $y$-axis scales, that the error stops decreasing with cell size for coarse RBF networks regardless of the degree of elements used. The representation error dominates and prevents the total error from decreasing, especially for $\norm{e_p}_{\mathcal{L}2}$.

\subsection{Adaptive RBF}

We verify that using adaptive refinement when producing the background grid during RBF generation greatly decreases the memory and training cost while preserving accuracy. To quantify this effect, adaptively refined RBF are compared to uniformly refined RBF. 
We compare results obtained using both types of grids with an equal cell size at the solid boundary. The Lethe background grid is composed of $64^3$ elements. RBF-generation, based on the convex superquadric (middle shape in Figure \ref{fig:stl_png_all_shapes}),
uses a base grid of 
$(2^{7-n})^3$
cells, where $n$ is the number of adaptive refinements.
The grid is subsequently adaptively refined $n$ times before the generation of RBF-nodes. This implies that, for all numbers of adaptive refinements tested, the distance between RBF-nodes close to the solid surface is the same.

\begin{figure}[htpb!]
	\centering
	\includegraphics[width=0.6\textwidth]{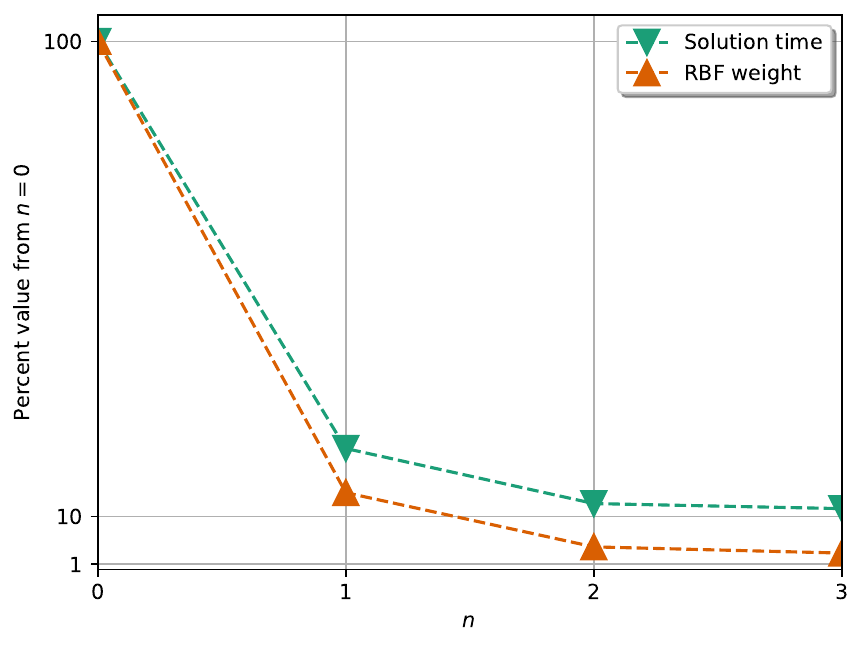}
	\caption{Relative value for solving time and RBF memory requirement compared to when only uniform refinement is used.} The grid size before adaptive refinement is $(2^{7-n})^3$. The shape is the convex superquadric.
	\label{fig:mms_adaptive}
\end{figure}

\textbf{Figure \ref{fig:mms_adaptive}} shows that decreasing the density of RBF-nodes far from the surface substantially decreases the solution time and memory requirements of the RBF-encoding, by factors 10 and 100, respectively. After two cycles of adaptive refinement of the RBF-producing background grid, the costs stop decreasing, which suggests that while this strategy is advantageous, its effects reach a plateau value. Specifically, the plateau is reached because almost all initial cells get refined at least once when $n=3$, since they are so large that they overlap with the shape. The first adaptive refinement cycle is therefore quasi-uniform. Increasing the padding of the bounding box could delay the plateau to higher values of $n$, but this approach would not reduce the absolute solution time or memory requirements.

\section{Validation in porous monoliths}
\label{sec:results}
    The SIBM implementation in Lethe has been validated by Barbeau et al. \cite{barbeau2024} for the coupled flow around moving spherical particles: Ten Cate sedimentation case \cite{ten2002validationSedim} ($\textit{Re} \leq 31.9$), positively buoyant sphere ($\textit{Re}=2500$), pair of drafting, kissing and tumbling (DKT) particles \cite{glowinski2001fictitious} ($\textit{Re} = 233$), and sedimentation of 64 particles ($\textit{Re} = 20$). We focus on validating the flow through monoliths at Reynolds number values under $30$.
    
    \subsection{Experimental pressure drop evaluation setup\label{sec:validation_exp}}
\begin{figure}[htpb!]
        \centering
        \includegraphics[width=0.9\textwidth]{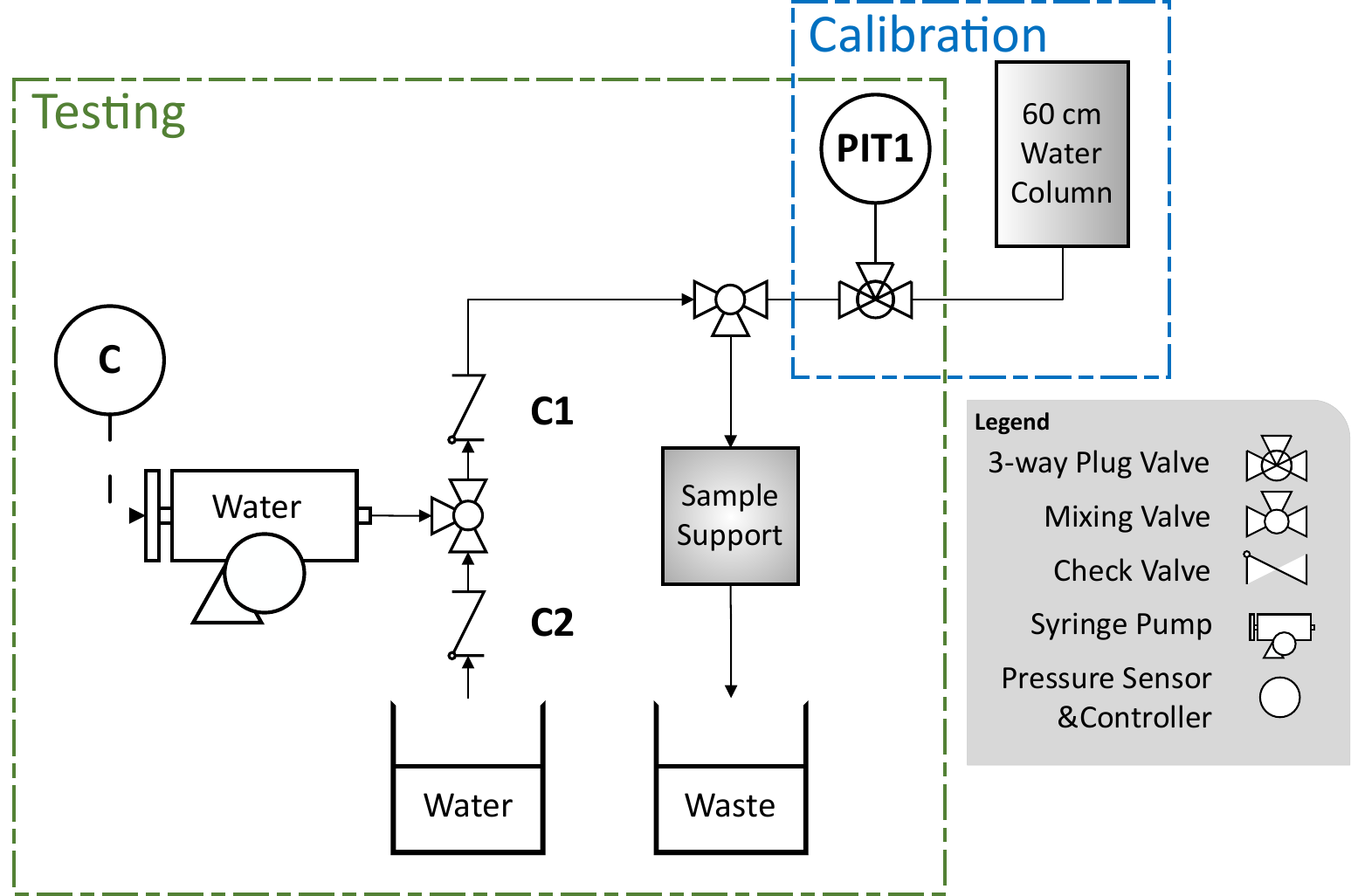}
        \caption{Schematic of the experimental setup used to automatically impose set volumetric flows to measure pressure drops in porous monolith samples.}
        \label{fig:schema_montage}
    \end{figure}

    We measure the pressure drop at different volumetric flows ($Q=\{0.6,0.9,1.2,1.5,1.8\}$ mL/s) through three (3) supported porous monolith samples using a controlled syringe pump. 
    This DIY syringe pump (RobotDigg, Shanghai, China) is connected to the support with an assembly of two check valves ($0.5$ psi break pressure) and a mixing valve that allow the pump to automatically refill itself, allowing for fewer manipulations and easier repetitions.
    During the filling phase, water flows from the feed reservoir into the syringe, while the check valve between the mixing valve and the monolith (C1) remains closed due to negative relative pressure. When the syringe is emptied, water is directed to the monolith as C1 is then subjected to positive pressure; at the same time, the check valve between the syringe and the feed reservoir (C2) is blocked. The valves are pressure-actuated.
    A differential pressure sensor (MPX5010DP from NXP, 0-10 kPa range) and a cylinder filled with water are also connected for pressure drop measurements and sensor calibration. The pressure sensor is connected to the tubing network on one side and open to the atmosphere on the other. Sensor calibration is done using a water column when the 3-way plug valve is correctly positioned. The samples are fit in the support using latex tubing having an inner diameter of  8 mm. \textbf{Figure \ref{fig:schema_montage}} shows the elements composing the experimental setup and a photograph of the setup is provided in \textbf{\ref{sec:appendix_experimental_setup}}. We use distilled water at a temperature of $24$ $^{\circ}$C.
    
    We calibrate the pressure sensor in an initial step during which the 3-way plug valve connects a  600 mm water column to the sensor. 
    Before testing each sample, we execute an empty run with the same sequence of volumetric flows. We subtract the values of the pressure drop when the sample support is empty from the measured pressure drop to isolate the porous monoliths contribution, which are to be compared with simulation results. The data for empty runs are shown in \textbf{\ref{sec:appendix_blank}}. In addition to storing monoliths in distilled water, we run cycles of low and high flow rates prior to measurements to ensure that any residual air is removed and that the flow is monophasic.
    
    \subsection{Simulation setup\label{sec:sim_setup}}
\begin{figure}[htpb!]
	\centering
	\includegraphics[width=0.8\textwidth]{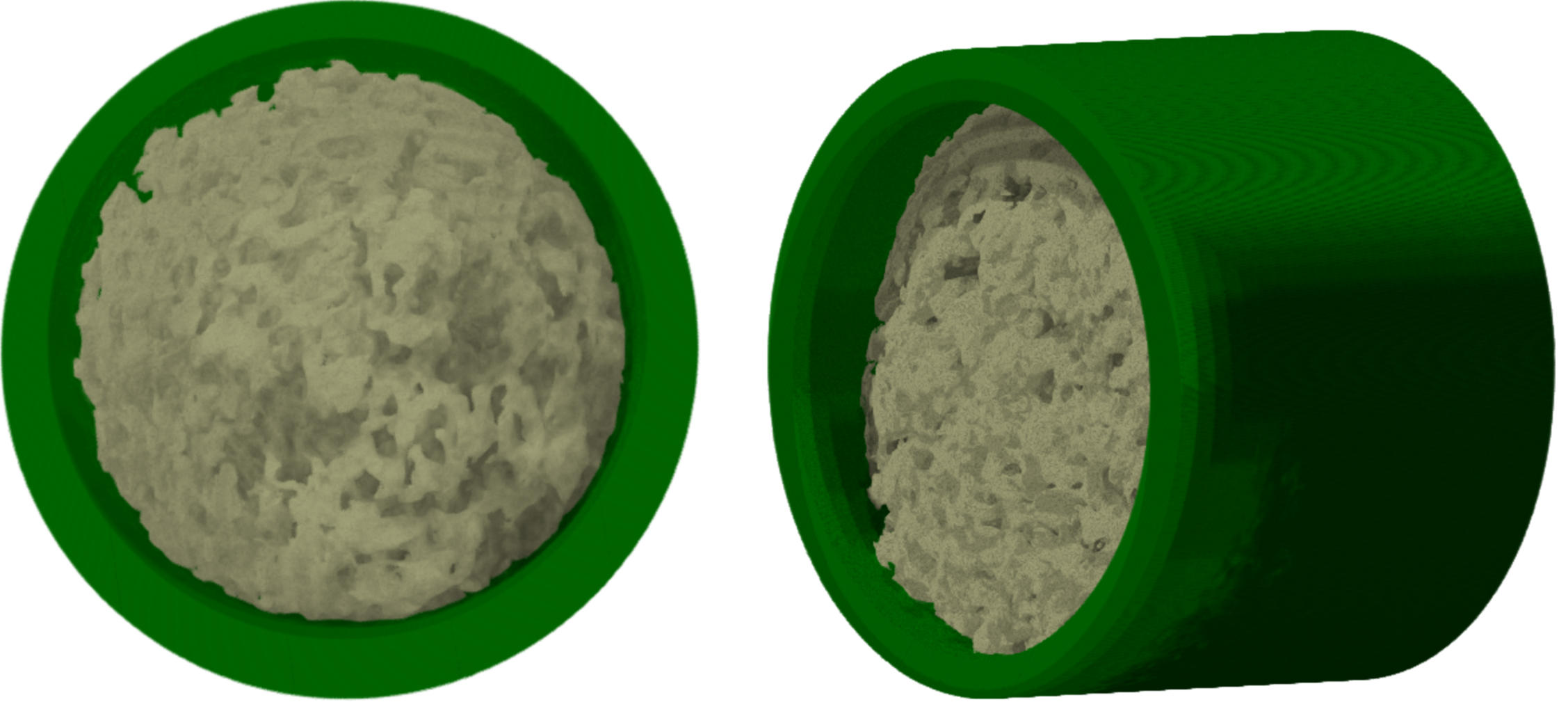}
	\caption{3D views (front and side) of a porous monolith and surrounding sleeve, exported from Dragonfly.}
	\label{fig:screenshot_dragonfly}
    \end{figure}

    We validate the methodology by comparing the simulated and measured pressure drops of each sample. For each sample, we generate 
    representations
    at various thresholds $\{1, 50, 99\}$ \% (low extreme, midpoint, high extreme), based on the parameter explained in Section \ref{sec:section_tomo}. We then use the RBF-encoding tool with the parameters shown in \textbf{Table \ref{tab:rbf_generation_parameters}} to construct RBF networks for both porous monoliths and the latex sleeve surrounding samples, to ensure that the domain is properly represented.
    \textbf{Figure \ref{fig:screenshot_dragonfly}} shows 3D views of the setup as exported from Dragonfly. The sleeve is tightly fitted around the sample to avoid any bypass of the flow, and since both the sleeve and sample are flexible and deformable they are digitized together.

    \begin{table}[htpb!]
            \centering
        	\caption{RBF encoding parameters.}
        	\label{tab:rbf_generation_parameters}
        	\begin{tabular}{c c }
        		\hline
        		\textbf{Parameter} & \textbf{Value} \\
        		\hline
        		\hline
        		  $m$           & $6$ \\
        		  $n$           & $2$ \\
        		  $\alpha$      & $3$ \\
        		  $p$           & $10$ \\
        		  Basis function    & C2 Wendland \\
        		\hline
        	\end{tabular}
        \end{table}

    We set the CFD parameters used in Lethe as shown in \textbf{Table \ref{tab:lethe_sharp_parameters}}; length unit is centimeter, time unit is second and mass unit is gram.
    We run the simulations in transient mode until a steady state is reached because the stiffness of the problem, the complex geometry, and the fact that the initial solution is too far from the stationary solution prevent the steady-state solver from converging.
    The elements are of degree 1 (second order accurate) for both velocity and pressure (Q$_1$Q$_1$). The inlet boundary conditions are set for velocity ($\bm{u}=[u_x,u_y,u_z]$) as Dirichlet, as $u_x=\{0.6,0.9,1.2,1.5,1.8\}$, $u_y=0$ and $u_z=0$. The side walls boundary conditions are set to $\bm{u}=\bm{0}$. The outlet boundary is set to $\nu \nabla \mathbf{u} \cdot \mathbf{n} - p^* \mathcal{I} \cdot \mathbf{n} = \bm{0}$, 
    where 
    $\bm{n}$ is the boundary normal vector, and $\mathcal{I}$ is the identity matrix. 
    The initial solution is $u_x=\{0.6,0.9,1.2,1.5,1.8\}$, $u_y=0$, $u_z=0$ and $p^*=0$.
    The background grid is a hyperrectangle defined by the diagonal $[-1.5,-0.5,-0.5]:[1.5,0.5,0.5]$ and consists of only cubic cells.

    \begin{table}[htpb!]
        \centering
        \caption{CFD parameters for the flow through porous media, where time units are seconds and length units are centimeters.}
        \label{tab:lethe_sharp_parameters}
        \begin{tabular}{c c }
            \toprule
            \textbf{Parameter} & \textbf{Value} \\
            \cmidrule(lr){1-2}
              Time integration scheme & BDF1 \\
              Time step               & $1\text{e}^{-4}$ \\ 
              Time end                & $2\text{e}^{-3}$ \\
              Density     &  $1$ \\
              Kinematic viscosity     & $1\text{e}^{-2}$ \\ 
              Element type            & Q$_1$Q$_1$             \\ 
              Initial solution        & $u_x=\{0.6,0.9,1.2,1.5,1.8\}$\\
              Inlet boundary          & $u_x=\{0.6,0.9,1.2,1.5,1.8\}$\\
              Side wall boundary      & $\bm{u}=\bm{0}$  \\
              Outlet boundary         & $\nu \nabla \mathbf{u} \cdot \mathbf{n} - p^* \mathcal{I} \cdot \mathbf{n} = \bm{0}$  \\
              Domain dimensions       & $[3,1,1]$       \\
            \bottomrule
        \end{tabular}
    \end{table}

    \subsection{Grid sensitivity analysis based on pressure drop \label{sec:grid_convergence}}
    
    We use h-adaptive refinement at fluid-solid interface to 
    generate the grids used for validation.
    Given the multiple inlet velocities and thresholds tested, we produce a grid convergence study for each extremum combination starting from a cartesian grid of $[3,1,1] \times 2^6$ cells in $[x,y,z]$ directions, adaptively refined thrice.
    We use a three point Richardson extrapolation based on the Q$_1$Q$_1$ interpolation's second order of convergence and compute the grid convergence indices (GCI) using a security factor (FS) of $1.25$ \cite{oberkampf2010vv,roache1998gci}. Grids are systematically refined between each level presented, but remain non-uniform.
    We consider only sample \#1, given its lowest pore size and under the assumption that the higher Reynolds numbers give the most mesh sensitive results.

    \begin{table}[htpb!]
        \centering
        \caption{Simulated pressure drops and GCI at various thresholds, flows and cell sizes, with a three point Richardson extrapolation ($\text{order of convergence}=2$).}
        \label{tab:convergence_study}
        \begin{tabular}{c c c c c c c c c c}
            \toprule
            
            \multirow{1}{*}{\bfseries Inlet velocity }[cm/s] & 
            \multicolumn{3}{c}{ 0.6}& 
            \multicolumn{3}{c}{ 1.2} & 
            \multicolumn{3}{c}{ 1.8} \\ 
            \cmidrule(lr){2-4}
            \cmidrule(lr){5-7}
            \cmidrule(lr){8-10}
            \textbf{Threshold} [\%] & 1 & 50 & 99 & 1 & 50 & 99 & 1 & 50 & 99 
            \\ 
            \hline
            \textbf{Smallest cell size } [$\mu$m]
            & \multicolumn{9}{c}{\textbf{$\Delta p$} [kPa]}
            \\
            \cmidrule(lr){1-10}
            19.53 & 0.4379 & 0.3398 & 0.1745 & 1.0697 & 0.7954 & 0.3766 & 1.9701 & 1.4245 & 0.6305 \\
            13.03 & 0.3807 & 0.3115 & 0.1640 & 0.8760 & 0.6793 & 0.3494 & 1.5347 & 1.1532 & 0.5720 \\
            9.77 & 0.3632 & 0.2925 & 0.1607 & 0.8221 & 0.6481 & 0.3412 & 1.4143 & 1.0952 & 0.5551 \\

            \cmidrule(lr){1-10}

            \textbf{Richardson extrapolation}
            & 0.3490 & 0.2733 & 0.1580 & 0.7803 & 0.6244 & 0.3346 & 1.3213 & 1.0572 & 0.5417 \\
            
            \cmidrule(lr){1-10}

            \textbf{Cell size } [$\mu$m]
            & \multicolumn{9}{c} {GCI [\%]}
            \\
            \cmidrule(lr){1-10}
            19.53 & - & - & - & - & - & - & - & - & - \\
            13.02 & 15.02 & 9.09 & 6.40 & 22.11 & 17.09 & 7.78 & 28.37 & 23.53 & 10.23 \\
            9.77 & 7.74 & 10.44 & 3.30 & 10.54 & 7.74 & 3.86 & 13.68 & 8.51 & 4.89 \\
            \bottomrule
        \end{tabular}
    \end{table}

\begin{figure}[hbtp!]
    	\centering
    	\includegraphics[width=.7\textwidth]{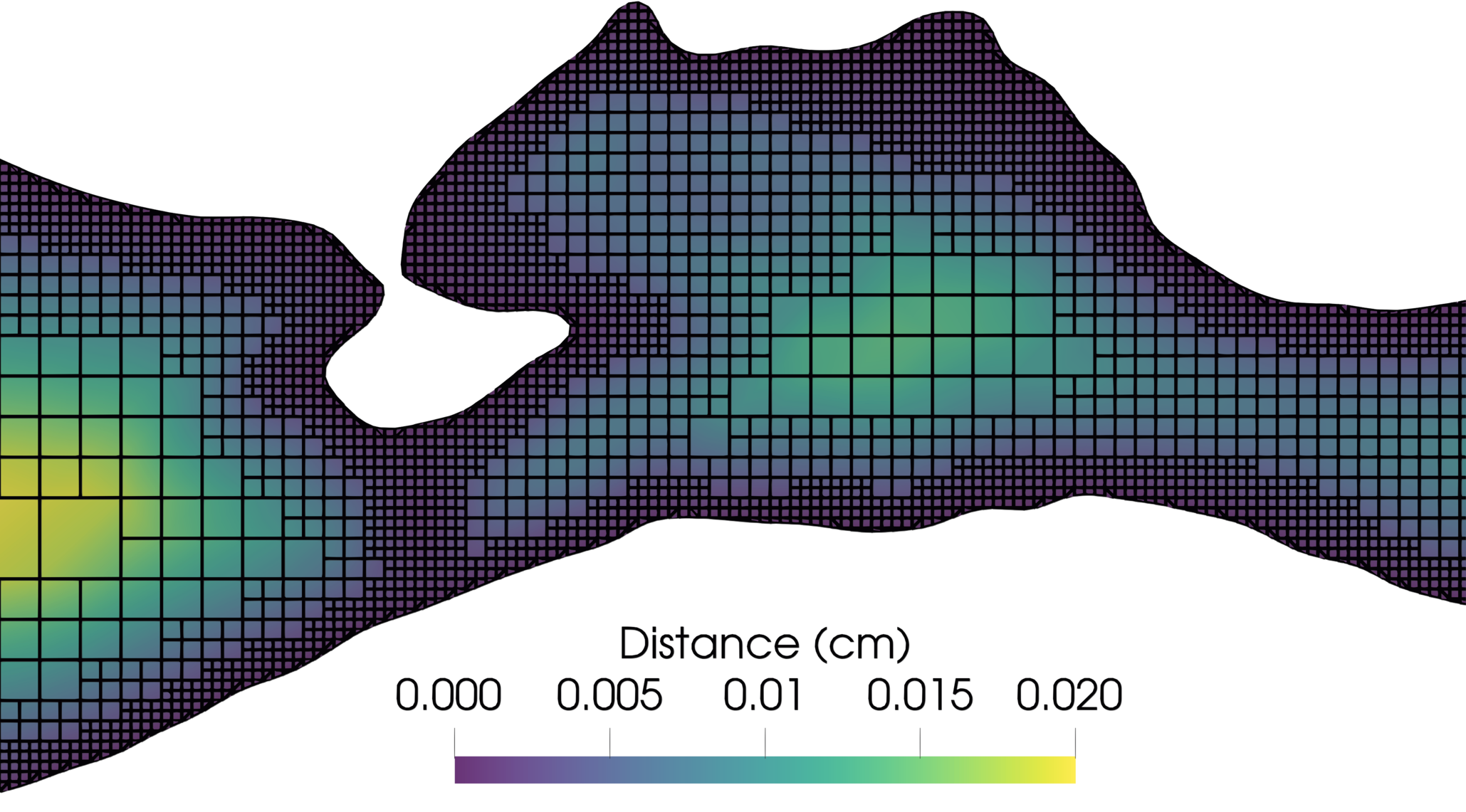}
    	\caption{Sliced view of a pore, with more refined cells near the solid.}
    	\label{fig:slice_pore_mesh}
    \end{figure}

    We conclude from \textbf{Table \ref{tab:convergence_study}} that using a cell size of 9.77 $\mu$m at the solid surface produces results with a discretization error of at most 14 \% for pressure drop predictions. The combination of a threshold of $1$ \% and an inlet velocity of $0.6$ cm/s results in the highest discretization error, which is of the magnitude of variations induced by the selection of the threshold parameter, and as such is considered acceptable. 
    \textbf{Figure \ref{fig:slice_pore_mesh}} shows a sliced view of a pore, with more refined cells near the solid surface, which is a result of distance-based adaptive refinement. For completeness, we included a similar 3D view in \textbf{\ref{sec:appendix_adaptive_3d}}.
    We reduce the number of required cells by a factor 16 by using adaptive mesh refinement.
    The final grids therefore have cell sizes between $9.77$ and $78.13$ $\mu$m; the total number of cells reaches $200$M and the simulations are run using 8000 cores
    spread across 200 compute nodes from the Digital Research Alliance of Canada's cluster Niagara.
    Each of Niagara's node uses 2 sockets with 20 Intel Skylake cores (2.4 GHz, AVX512), for a total of 40 cores per node. The network connection is a 100 Gb/s EDR Dragonfly+. Each node has access to 202 GB of Random Access Memory.

    \subsection{Validation with silicone porous monoliths}    
\begin{figure}[htpb!]
    	\centering
    	\includegraphics[width=\textwidth]{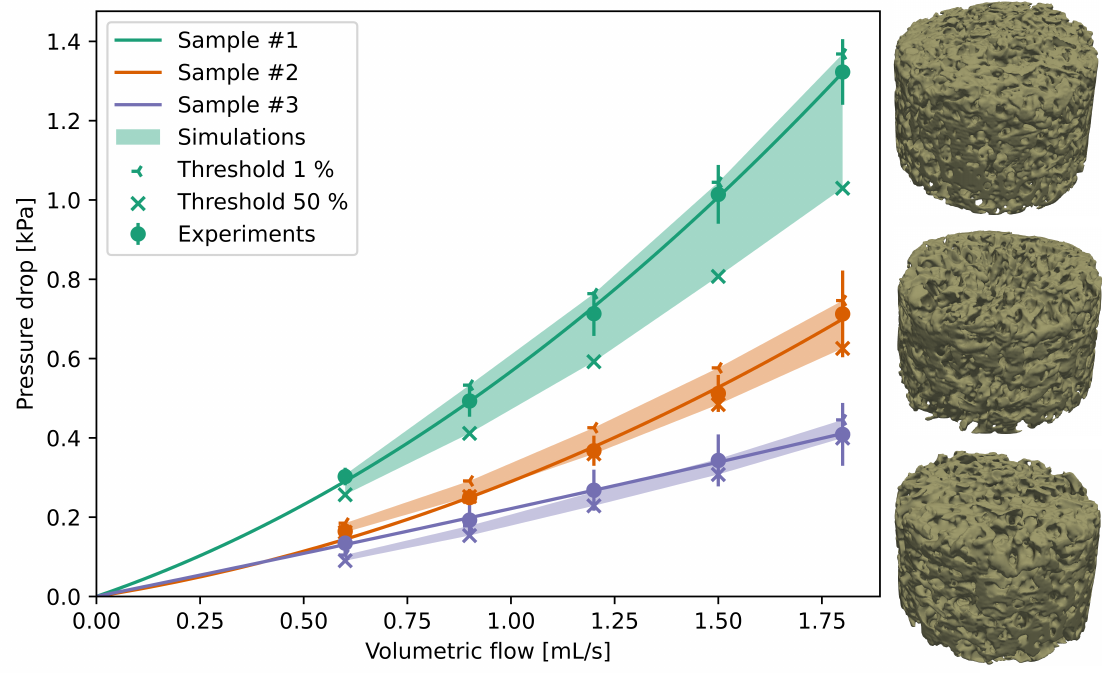}
    	\caption{Experimental and simulated pressure drops of three silicone porous monoliths (average pore size: 300 $\mu$m). Darkened zones represent the values covered by thresholds between 1 \% and 50 \%. Experimental error bars are the T-Student's confidence interval using $\alpha = 0.05$ (n = 5 number of repetitions). Each porous monolith (right) represents a sample and is aligned to its corresponding curve.}
    	\label{fig:validation_pressure_drop}
    \end{figure}

    We compare the simulation results of each porous monolith, with a range of applied flow rates $Q=[0.6,1.8]$ mL/s and with surface meshes generated using different thresholds $\{1,50,99\}$ \%. 
    We discard simulated pressure drops at 99 \% threshold because these values are unrealistically low and do not align with experimental observations (see \textbf{\ref{sec:appendix_validation_threshold99}}).    
    We show the results for each sample in \textbf{Figure \ref{fig:validation_pressure_drop}}. The experimental error bars correspond to the confidence intervals using a T-Student distribution, with a significance level $\alpha=0.05$.    
    The agreement is slightly worse at low flow rates for samples \#2 and \#3, but still acceptable. 
    We attribute this difference to the low signal obtained from the pressure sensor below $0.2$ kPa and the low accuracy of the system for small flow rates.
    The experimental error is of the same magnitude as the error induced by the selection of the threshold parameter, which confirms that the cell size selected in Section \ref{sec:grid_convergence} is appropriate. None of the three main sources of error for validation (discretization, threshold and pressure measurements) stand out or would be worth reducing. 
    
\section{Analysis}
\label{sec:analysis}

We use the CFD results of the flow through each sample to gain insight on the phenomena at the microscopic level. The pressure field is first observed slice by slice along the flow axis, both in terms of mean and standard deviation. 
Then, we highlight the impact of the 
principal pathways (pore and channels networks)
on the progression of pressure fronts and local Reynolds numbers through a sample.

\subsection{Pressure drop along the flow axis}
\begin{figure}[htpb!]
    	\centering    	\includegraphics[width=\textwidth]{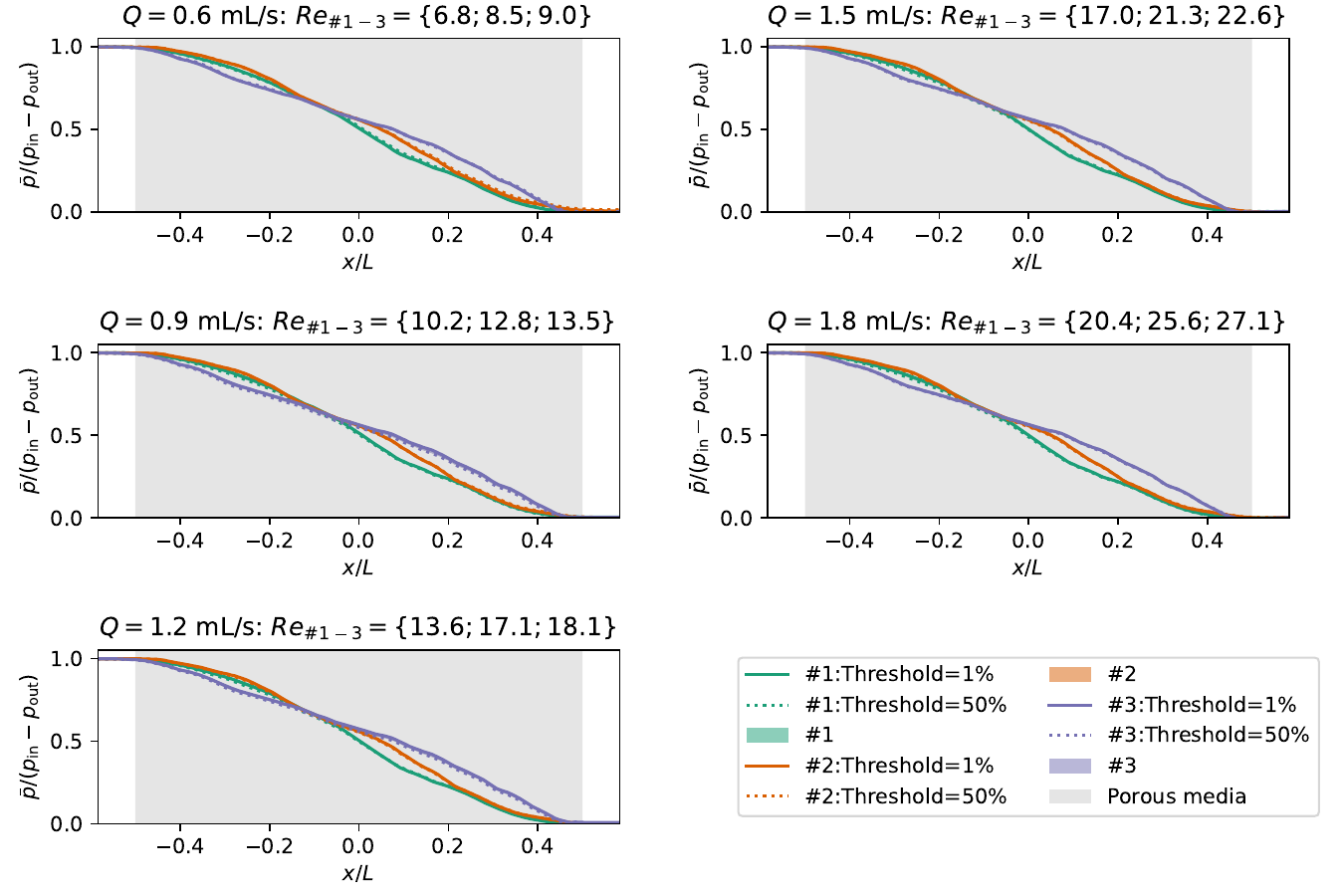}
    	\caption{
                Mean pressure evolution along the flow axis. Each subplot represents a set volumetric flow rate. Reynolds numbers vary for a given flow rate due to specific monolith properties.}
    	\label{fig:pressure_drop_along_axis}
    \end{figure}

\begin{figure}[htpb!]
    	\centering
    	\includegraphics[width=\textwidth]{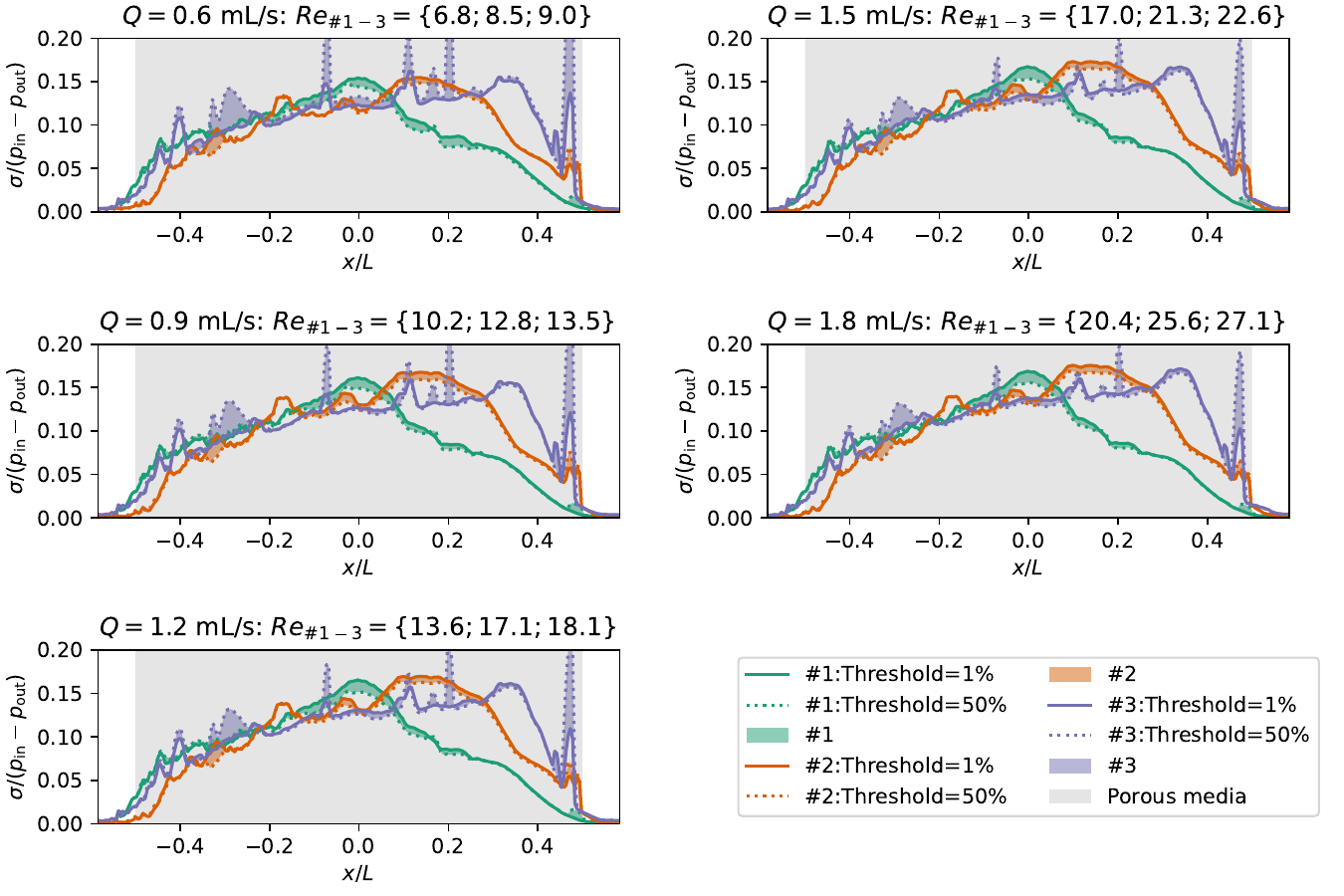}
    	\caption{
                Standard deviation ($\sigma$) of pressure along the flow axis. Each subplot represents a set volumetric flow rate. Reynolds numbers vary for a given flow rate due to specific monolith properties.}
    	\label{fig:std_pressure_along_axis}
    \end{figure}

    The mean pressure $\bar{p}$ is taken as surface average of the fluid phase of a given slice at axial position $x$, as described by \textbf{Equation \ref{eq:slice_mean}}, where $\theta$ is the variable ($p$ here). 
    The standard deviation at each axial position is described by \textbf{Equation \ref{eq:slice_std}}.
    The mean pressure drop along the axis is shown in \textbf{Figure \ref{fig:pressure_drop_along_axis}}, as well as the standard deviation of the pressure in \textbf{Figure \ref{fig:std_pressure_along_axis}}. The Reynolds number is defined as the porous Reynolds number in \textbf{Equation \ref{eq:reynolds}} and the Darcy velocity ($u_\text{Darcy}$) is defined in \textbf{Equation \ref{eq:darcy_velocity}}, computed with volumetric flow $Q$, cross-sectional area $A_\text{section}$, and porosity \cite{wood2020review_turbFlo_porousmedia}.

    \begin{equation}
        \bar{\theta}(x) = \left( \frac{\iint_{\Omega_\text{fluid}} \theta d\Omega} {\iint_{\Omega_\text{fluid}} d\Omega} \right)_x
        \label{eq:slice_mean}
    \end{equation}
    
    \begin{equation}
        \sigma_{\theta}(x) = \left( \frac{\iint_{\Omega_\text{fluid}} \left(\theta  -  \bar{\theta}(x)
        \right)^2 d\Omega} {\iint_{\Omega_\text{fluid}} d\Omega} \right)^{1/2}_x
        \label{eq:slice_std}
    \end{equation}

    \begin{equation}
        \textit{Re} = \frac{u_\text{Darcy} d_\text{solid}}{\nu }
        = \frac{Q d_\text{solid}}{A_\text{section} \varepsilon \nu }
        \label{eq:reynolds}
    \end{equation}

    \begin{equation}
        u_\text{Darcy}=\frac{Q}{A_\text{section} \varepsilon}
        \label{eq:darcy_velocity}
    \end{equation}


Mean pressure evolutions (Figure \ref{fig:pressure_drop_along_axis}) indicate that, at this scale, the topology of the porous networks act as the principal factor to describe the pressure evolution through the monoliths: the threshold parameter and flow rate affect the pressure by a factor, but not the shape of the curves.
This makes sense given that all flow rates tested correspond to a laminar regime, with $\textit{Re} \in [6.8,27.1]$.
The differences between the normalized curves are therefore entirely due to the unique structure of each monolith.
In agreement with these observations, the pressure standard deviations (Figure \ref{fig:std_pressure_along_axis}) follow the same patterns for each simulated rate, exhibiting sharp variations at the same locations.
The only exception to the observation regarding the threshold is the presence of localized peaks for sample \#3 in Figure \ref{fig:std_pressure_along_axis}. 
Sample \#3 with a threshold of 50 \% exhibits standard deviation peaks that slightly attenuate at higher flow rates. A hypothesis for this phenomenon is that some smaller channels are blocked when the threshold is 1 \%, while these same channels are open for a threshold of 50 \%. Observing this phenomenon only for sample \#3 would be due to its high standard deviation for the pore diameter: $132$ $\mu$m ($+42$ \% higher than others).

\begin{figure}[htpb!]
    	\centering
    	\includegraphics[width=\textwidth]{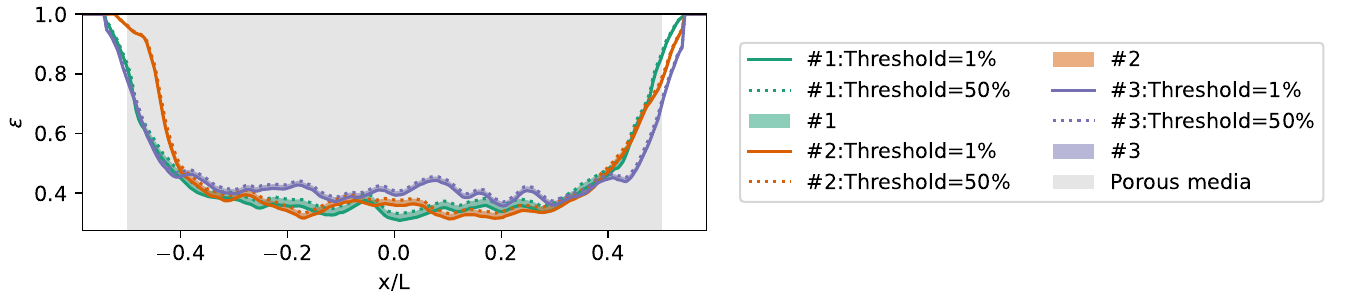}
    	\caption{Local porosity of the porous monoliths along the flow axis, for thresholds of 1 and 50 \%.}
    	\label{fig:local_poro_simple}
    \end{figure}

In addition, entry and exit regions are apparent and represent the major part of the domain; the slope of $\bar{p}$ varies throughout the monoliths and tends to zero at the edges. This is due to the varying compression level of the pores depending on the axial position. As seen in \textbf{Figure \ref{fig:local_poro_simple}}:  the regions closer to edges (entry and exit zones) present higher porosity, going from about $0.4$ (center of the domain) to $1$ (edges). This effect is possibly due to the synthesis process, but also to the way the monoliths are pushed into the support in a tight manner to avoid bypassing. The monoliths are therefore compressed more near the center of their axial position while the pores at the edges remain more open. 
The resulting effect is that regions at the edge are more porous, resulting in pressure gradients closer to zero.

\subsection{Preferential channels and pressure fronts}
\begin{figure}[htpb!]
	\centering
	\includegraphics[width=\textwidth]{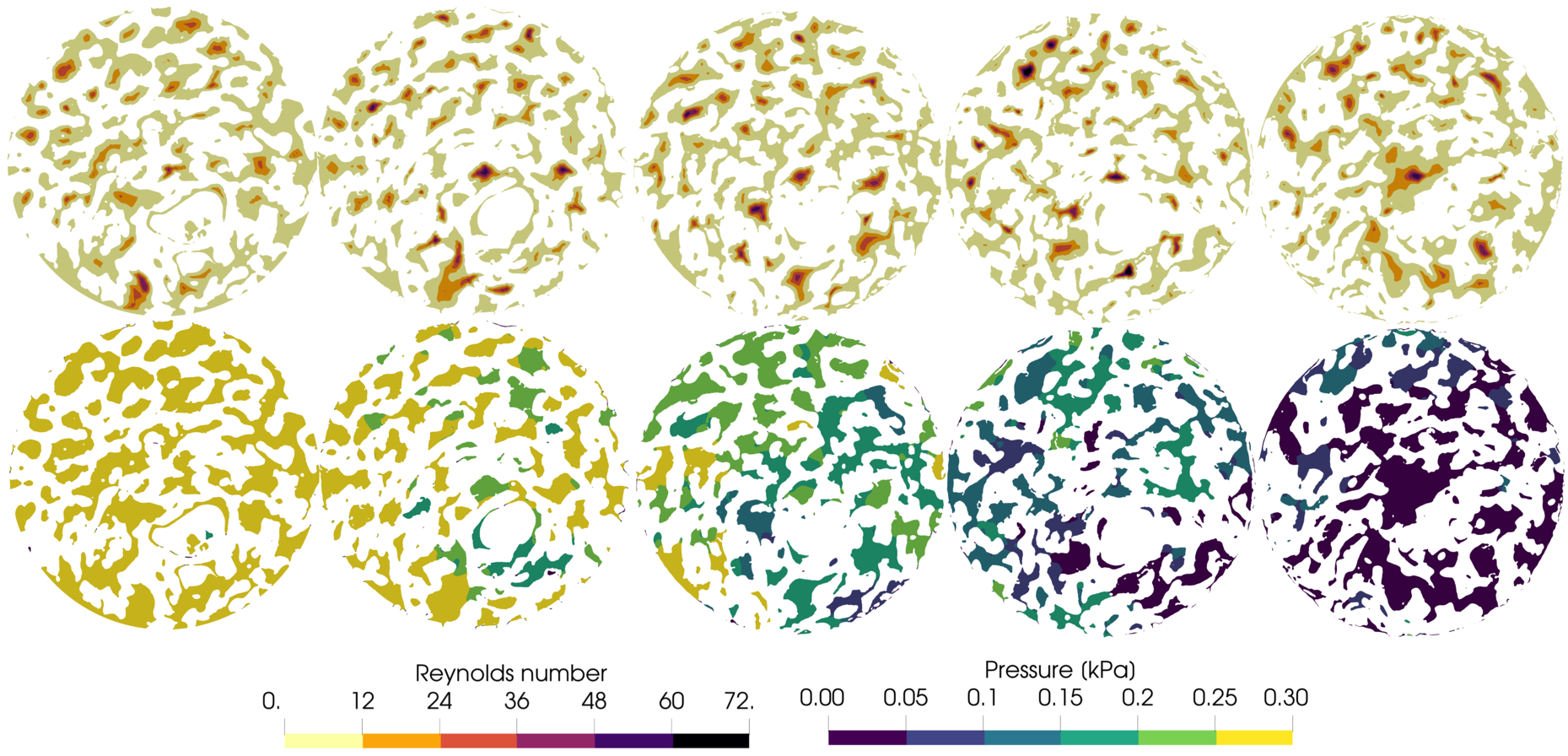}
	\caption{Local Reynolds number (top row) and pressure field (bottom row) evolving along the flow axis for sample \#2, at a flow rate of $1.2$ mL/s, with a threshold of 50 \%. The slices are taken at $x=\{-0.2,-0.1,0,0.1,0.2\}$ cm.}
	\label{fig:montage_fronts_transparent}
\end{figure}

The porous silicone monoliths exhibit heterogeneities: different pore scales, structures appearing over multiple slices and a few principal channels.
Because of the variability of the channels structure and their high impact on the pressure field, we can expect the fluid elements to progress along the flow axis at varying rates, which can be seen by the progression of pressure fronts and confirmed by local Reynolds numbers calculations.
\textbf{Figure \ref{fig:montage_fronts_transparent}} shows the locally computed Reynolds number ($\textit{Re}(\bm{x}) = \lambda(\bm{x})\norm{u}/\nu$) and pressure field through multiple slices for sample \#2, at a flow rate of $1.2$ mL/s, with a threshold of 50 \%. The slice at $x=0$ shows the magnitude of range of pressures that can be observed in a single slice, where variations can reach half the total pressure drop. More specifically, the pressure field at this position has both yellow and dark blue regions, which indicate a range of pressure of $0.2$ kPa.  

\begin{figure}[htpb!]
	\centering
	\includegraphics[width=.9\textwidth]{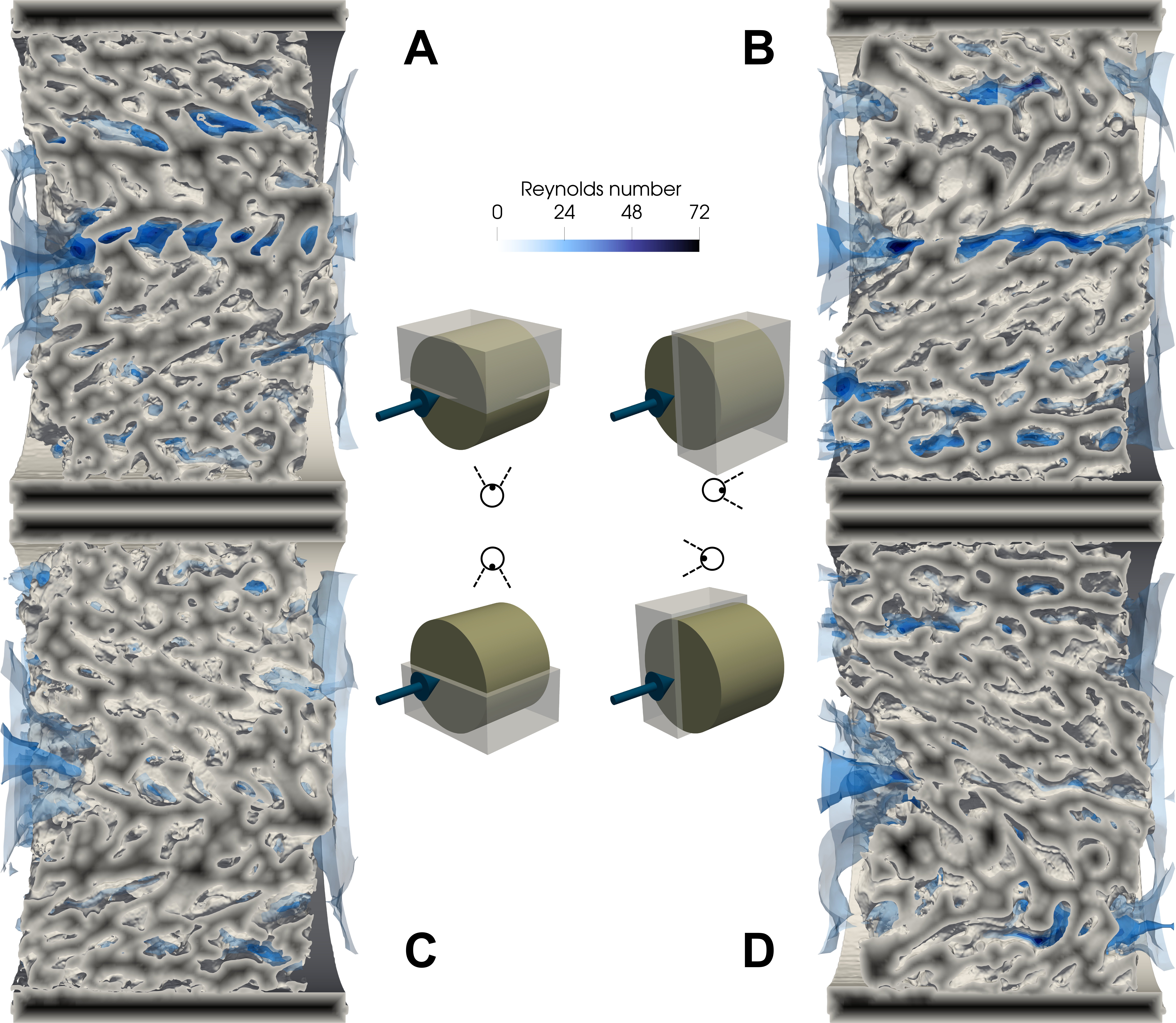}
	\caption{Reynolds number isovalues through sample \#2, at a flow rate of $1.2$ mL/s, with a threshold of 50 \%. Eye symbols represent the points of view of each half. The arrows represent the flow direction. }
	\label{fig:montage_reynolds_3d}
\end{figure}

\textbf{Figure \ref{fig:montage_reynolds_3d}} highlights that preferential channels go through the monolith with localized high isovalues of $\textit{Re}$. For example, a major preferential channel is visible in subfigures A and B at midheight, which coincides with the high Reynolds number peaks at the center of every slice of Figure \ref{fig:montage_fronts_transparent}. 
A few dead zones are also apparent and are in agreement with Figure \ref{fig:montage_fronts_transparent}. The regions surrounding the main central channel contain thick walls and pores only slightly connected to the pore network. In Figure \ref{fig:montage_reynolds_3d}, these regions are especially apparent in subfigures B and D in the second and third quartiles of the height. 

\section{Discussion}
\label{sec:discussion}

\subsection{Flow regimes}
 We choose the pore size mean as the characteristic length when computing Reynolds numbers. However, as seen from Table \ref{tab:samples_characterization}, pore size distributions of our samples are wide: large and small pores are present. \textbf{Table \ref{tab:pore_reynolds}} shows pore Reynolds numbers computed using extrema pore sizes (addition or subtraction of two standard deviations). For each sample, Table \ref{tab:pore_reynolds} highlights the sensitivity of the Reynolds number to this decision. Sample \#3, for example, shows variations up to a factor 6 for the Reynolds number between its extrema. 

     \begin{table}[htpb!]
        \centering
        \caption{Pore Reynolds numbers for each sample, for various volumetric flows ($Q=\{0.6,0.9,1.2,1.5,1.8\}$ mL/s) and using the pore sizes at the 2.28th, 50th and 97.72th percentiles. These percentiles correspond to the mean $\mu$ to which two standard deviations $\sigma$ are subtracted or added.}
        \label{tab:pore_reynolds}
        \begin{tabular}{c c c c c c c c c c}
            \toprule
            
            \multirow{1}{*}{\bfseries Sample \# } & 
            \multicolumn{3}{c}{ 1}& 
            \multicolumn{3}{c}{ 2} & 
            \multicolumn{3}{c}{ 3} \\ 
            \cmidrule(lr){2-4}
            \cmidrule(lr){5-7}
            \cmidrule(lr){8-10}
            \textbf{Characteristic length} & $\mu - 2 \sigma$ & $\mu$ & $\mu + 2 \sigma$ & $\mu - 2 \sigma$ & $\mu$ & $\mu + 2 \sigma$ & $\mu - 2 \sigma$ & $\mu$ & $\mu + 2 \sigma$
            \\ 
            \hline
            \textbf{Flow rate } [mL/s]
            & \multicolumn{9}{c}{\textbf{$\textit{Re}$}}
            \\
            \cmidrule(lr){1-10}
            0.6 &   2.2 & 6.8 & 11.4 & 4.6 & 8.5 & 12.5 & 2.5 & 9.0 & 15.5 \\
            0.9 &   3.3 & 10.2 & 17.1 & 6.9 & 12.8 & 18.7 & 3.8 & 13.5 & 23.3 \\
            1.2 &   4.4 & 13.6 & 22.8 & 9.2 & 17.1 & 25.0 & 5.0 & 18.1 & 31.1 \\
            1.5 &   5.5 & 17.0 & 28.5 & 11.7 & 21.3 & 31.2 & 6.3 & 22.6 & 38.8 \\
            1.8 &   6.7 & 20.4 & 34.2 & 13.8 & 25.6 & 37.4  & 7.5 & 27.1 & 46.6 \\
            
            \bottomrule
        \end{tabular}
    \end{table}

 The flow regimes are laminar, but they deviate from Darcy flow around the upper bounds of porous Reynolds number studied, while remaining steady flows \cite{dybbs1984}.

\subsection{Limitations}

     For a problem of the same scale (dimension, pore sizes), increasing the flow rate 
     would require finer grids to properly capture thinner boundary layers and, potentially, recirculation zones.
     Furthermore, it is unclear if such flow would be steady.
     A grid composed of cells twice finer than the ones set in Section \ref{sec:grid_convergence} would lead to a problem eight times larger. While challenging to solve, it would be possible with sufficient computing ressources. It would theoretically be possible to handle much finer discretizations using a matrix-free approach \cite{prieto2024matrix}, but this type of method is unavailable within our current implementation of the IBM. 
     
     In porous media of higher porosity and pore size (e.g. open-cell foams), it would be possible to handle problems in the transition regime. Limitations then would lie in the overall computing costs due to time integration. Caution would be needed however, to ensure that fine structures of the media are properly captured by the RBF networks. 

\section{Conclusion}
\label{sec:conclusion}

    Circumventing transport phenomena limitations is at the core of process scale-up, and understanding and predicting sub-millimetric flow characteristics is a key component to improving the performance of porous media based equipment. 
    In this work, we use pore-resolved CFD to accurately predict the flow through porous monoliths, providing a method for producing physically realistic digital twins of complex porous media. 

    RBF networks are generated from digitized monoliths and used in large scale simulations that cover complete centimeter-scale media with non-conformal cartesian grids, which alleviates the need for complex mesh generation. 
    The method is verified using the method of manufactured solutions, and validated against pressure drop measurements obtained experimentally.
    
    The proposed method is flexible in both usage and application fields, which makes it a powerful complementary tool for design as it provides understanding of the flow inside complex and possibly opaque media. 
    Media topology can come from digitization or the result of computer-aided design (CAD). Coupling PRCFD to additive      manufacturing could enable custom production of highly efficient components for heat exchangers, batteries and micro-porous reactors, paving the way for novel, efficient, and modular technologies.

\section{CRediT authorship contribution statement}
\label{sec:credit}
\textbf{Olivier Guévremont}: Conceptualization,  Formal analysis, Investigation, Methodology, Software, Validation, Visualization, Roles/Writing - original draft.
\textbf{Lucka Barbeau}: Conceptualization,  Methodology, Software, Writing - review \& editing. 
\textbf{Vaiana Moreau}: Methodology, Writing - original draft.
\textbf{Federico Galli}: Methodology, Supervision, Writing - review \& editing.
\textbf{Nick Virgilio}: Funding acquisition, Methodology, Resources, Supervision, Writing - review \& editing.
\textbf{Bruno Blais}: Conceptualization, Funding acquisition, Methodology, Resources, Software, Supervision, Writing - review \& editing.

\section{Declaration of competing interest}
\label{sec:declaration_of_interest}
The authors declare the following financial interests/personal relationships which may be considered as potential competing interests:
Bruno Blais reports financial support was provided by Natural Sciences and Engineering Research Council of Canada through the RGPIN2020-04510 Discovery Grant and the MMIAOW Canada Research Chair (level 2) in Computer-Assisted Design and Scale-up of Alternative Energy Vectors for
Sustainable Chemical Processes. 
Olivier Guévremont reports financial support was provided by Natural Sciences and Engineering Research Council of Canada, Fonds de recherche du Québec – Nature and technologies and Institut de l'Énergie Trottier.
Bruno Blais and Federico Galli report equipment, drugs, or supplies were provided by Digital Research Alliance of Canada.

\section{Data availability}
\label{sec:data_availability}
Data will be made available on request.

\section{Acknowledgments}
\label{sec:acknowledgments}

The authors would like to acknowledge the support of Lauriane Parès, for help in the laboratory, as well as Teodora Gancheva, for help in synthetizing the porous silicone samples and fitting them in the supports using flexible sleeves.
Olivier Guévremont would like to acknowledge the financial support of the Natural Sciences and Engineering Research Council of Canada (NSERC), the Fonds de recherche du Québec – Nature and technologies (FRQNT) and Institut de l'Énergie Trottier (IET).
The authors would also like to acknowledge the technical support and computing time provided by the Digital Research Alliance of Canada.

\bibliographystyle{elsarticle-num} 


\appendix
\section{Synthesis of porous silicone monoliths}
\label{sec:porous_samples_synthesis}
\begin{figure}[htpb!]
        \centering
        \includegraphics[width=\textwidth]{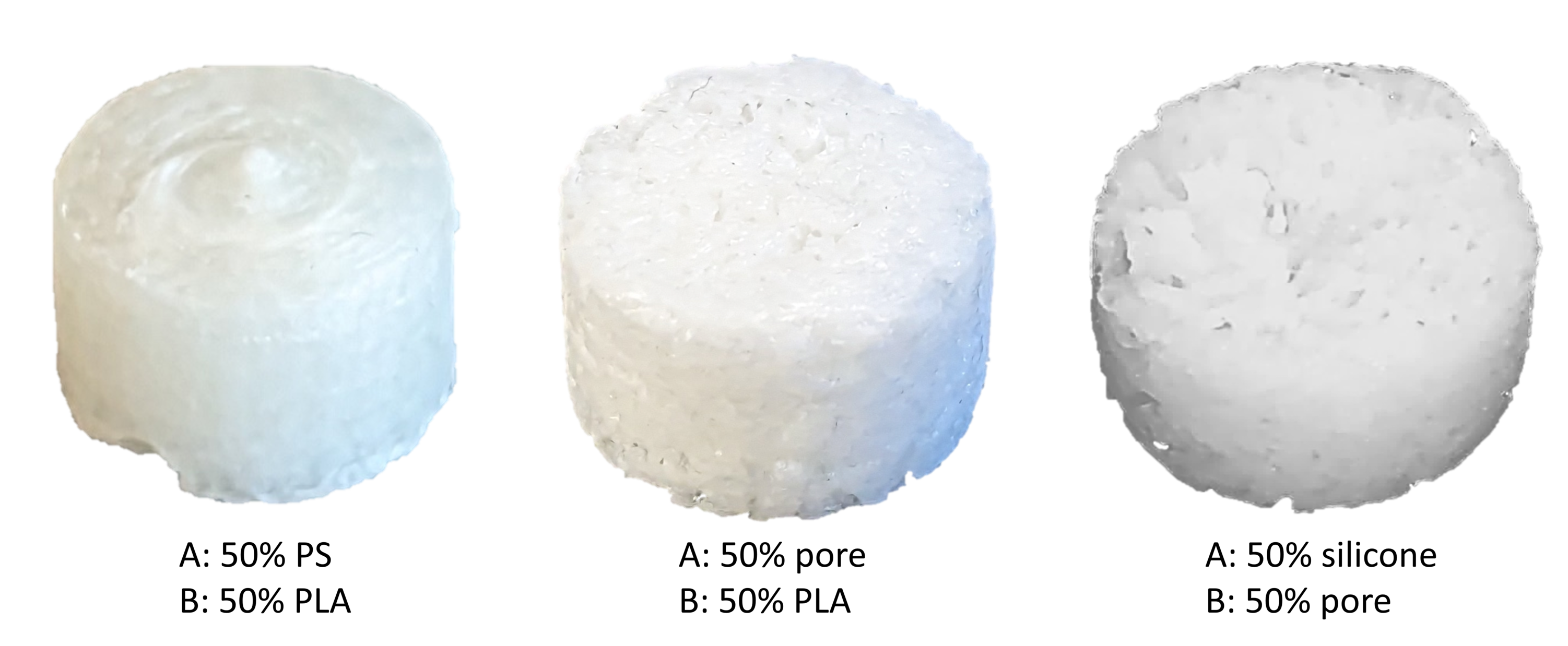}
        \caption{Fabrication of the silicone-based porous monoliths: PS/PLA 50/50 wt\% blend is extruded, PS is selectively extracted, then silicone is injected in the PLA mold after which the PLA is also removed.}
        \label{fig:experimental_workflow}
    \end{figure}

    We synthethize silicone porous samples by \textit{(1)} melt-processing of polystyrene (PS) and polylactic acide (PLA) blends, \textit{(2)} selective extraction of PS, and \textit{(3)} injection of silicone components in porous PLA molds, \textit{(4)} crosslinking and \textit{(5)} selective extraction of PLA. These steps are summarized in \textbf{Figure \ref{fig:experimental_workflow}}.

    \subsubsection{Preparation of porous PLA molds}    
    We prepare PLA porous molds 
    as previously described by Esquirol et al. \cite{esquirol2014cocontinuousPolymers}. First, we dry PS and PLA granules in a vacuum oven at 60 °C during 12 h. We then melt-process a blend of PS and PLA (50/50 wt\%) using an AG 34 mm Leistritz co-rotating twin-screw extruder at 190 °C. Following extrusion, we quench the rod-shaped blend in cold water to fix its morphology. We then anneal the blend under quiescent conditions with a hot press at 190 °C for 45 minutes, to let the morphology coarsen.
    After annealing, we quench the annealed polymer blend again in cold-water. We trim the annealed PS/PLA rods into cylinders (diameter $=0.8$ cm, and thickness $=0.6$ cm) by computer numerical control (CNC) machining. Next, we remove the PS phase by selective extraction in a Soxhlet extractor with cyclohexane for 1 week. After drying in a vacuum oven at 60°C for 2 days, we verify that all PS is extracted by gravimetry. We store the resulting porous PLA molds at room temperature until the next step. 

    \subsubsection{Preparation of porous silicone monoliths}
    We mix silicone precursor components
    (Shenzhen E4ulife Technology Co, product \#X002NKUNXF, Let's Resin silicone rubber) in a beaker and then transfer to 15 mL falcon tubes. We deposit porous PLA molds individually into the silicone-filled tubes and submerge them to ensure that the porous molds are entirely filled with the solution of silicone precursors. We deposit the tubes in a custom-built injection system applying vacuum/pressure cycles until air stops being released from the molds. We retrieve the silicone-loaded molds and place them in a petri dish at room temperature to allow complete gelation. After 24 h, we remove the excess silicone around the sides of the molds using a cutting tool. We then selectively extract the PLA phase using chloroform (Thermo ScientificTM \#42355, changed every two days) under agitation  for 10 days. We rinse the resulting silicone monoliths with MilliQ water to remove the residual chloroform. We finally dry the silicone monoliths in a vacuum oven at 60 °C for 2 h, weight the samples to verify PLA extraction, and store in distilled water at room temperature until use.  

\section{Porous characteristics distributions}
\label{sec:appendix_pore_size_distributions}

\begin{figure}[htpb!]
	\centering
    	\includegraphics[width=0.7\textwidth]{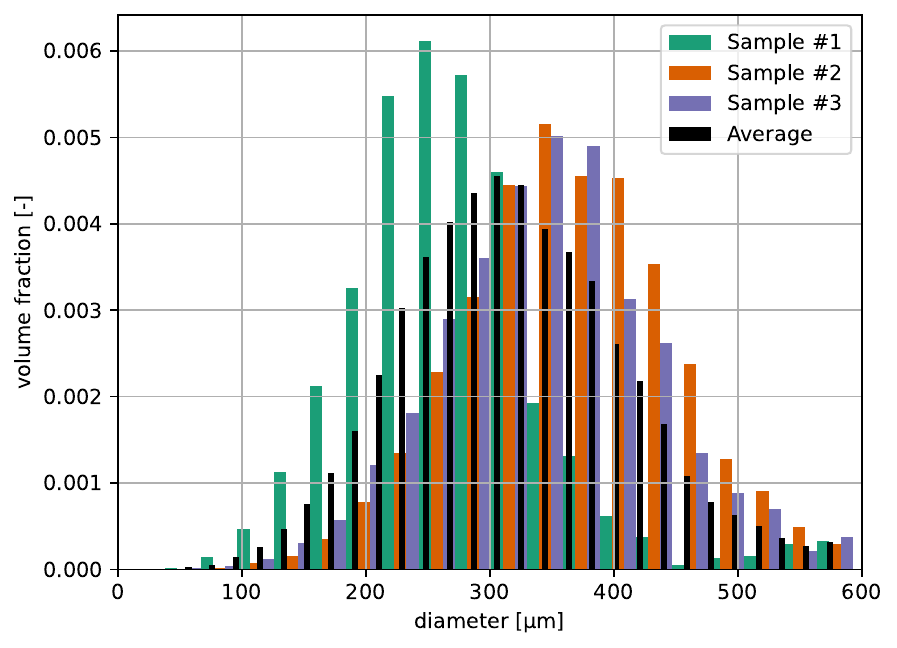}
	\caption{Volume-based pore size distribution of each sample.}
	\label{fig:pore_size_distribution}
    \end{figure}

    \textbf{Figure \ref{fig:pore_size_distribution}} shows the pore size distribution of each monoliths (\#1-3). The Gaussian shape confirms that describing the distributions in terms of mean and standard deviation is appropriate.

\begin{figure}[htpb!]
	\centering
    	\includegraphics[width=0.7\textwidth]{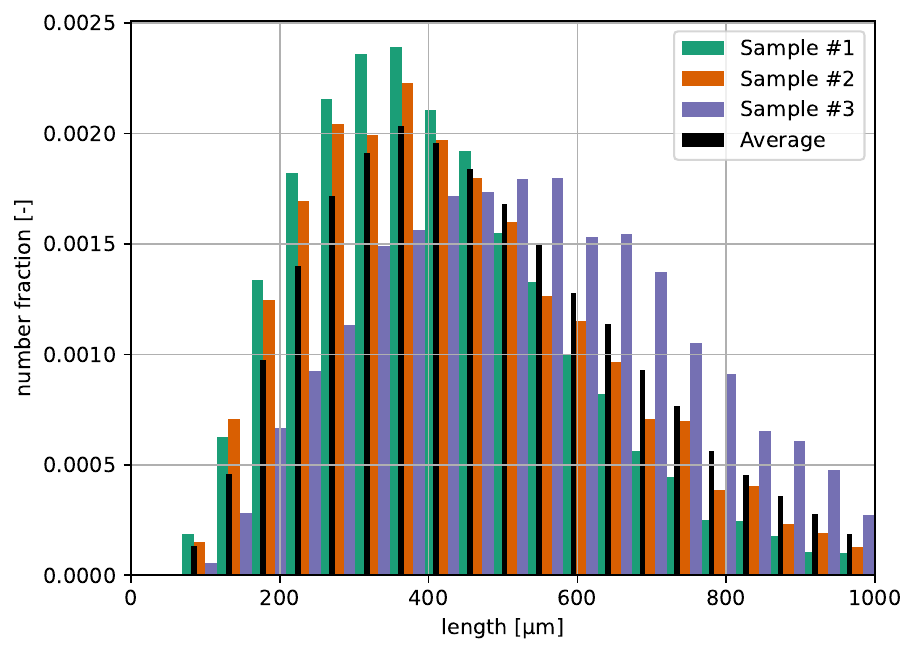}
	\caption{Number-based distance between pairs of connected pores, for each sample.}
	\label{fig:pore_distance_distribution}
    \end{figure}

    \textbf{Figure \ref{fig:pore_distance_distribution}} shows the distance between pairs of connected pores, for each monolith (\#1-3). The distributions are log-normal. The difference between the distributions of samples \#2 and \#3 in Figure \ref{fig:pore_distance_distribution} might explain the difference of their observed pressure drops.

\section{Experimental setup}
\label{sec:appendix_experimental_setup}
\begin{figure}[htpb!]
	\centering
	\includegraphics[width=0.5\textwidth]{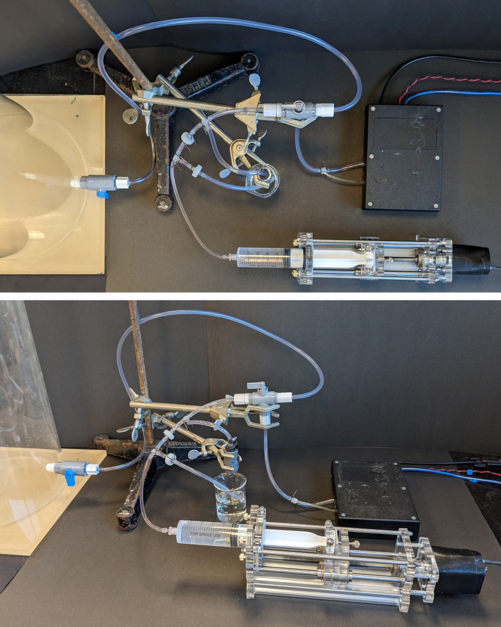}
	\caption{Photograph (top and side views) of the experimental setup. We can see the water column (left), the syringe pump (bottom), the controller (right) and the sample support (grey, center) above the beaker.}
	\label{fig:photograph_setup}
    \end{figure}

\textbf{ Figure \ref{fig:photograph_setup}} shows two pictures of the experimental setup used to measure the pressure drop as the fluid passes through the monolith at a given flow rate.

\section{Blank experimental data (without sample)}
\label{sec:appendix_blank}

\begin{figure}[htpb!]
        \centering
        \includegraphics[width=0.8\textwidth]{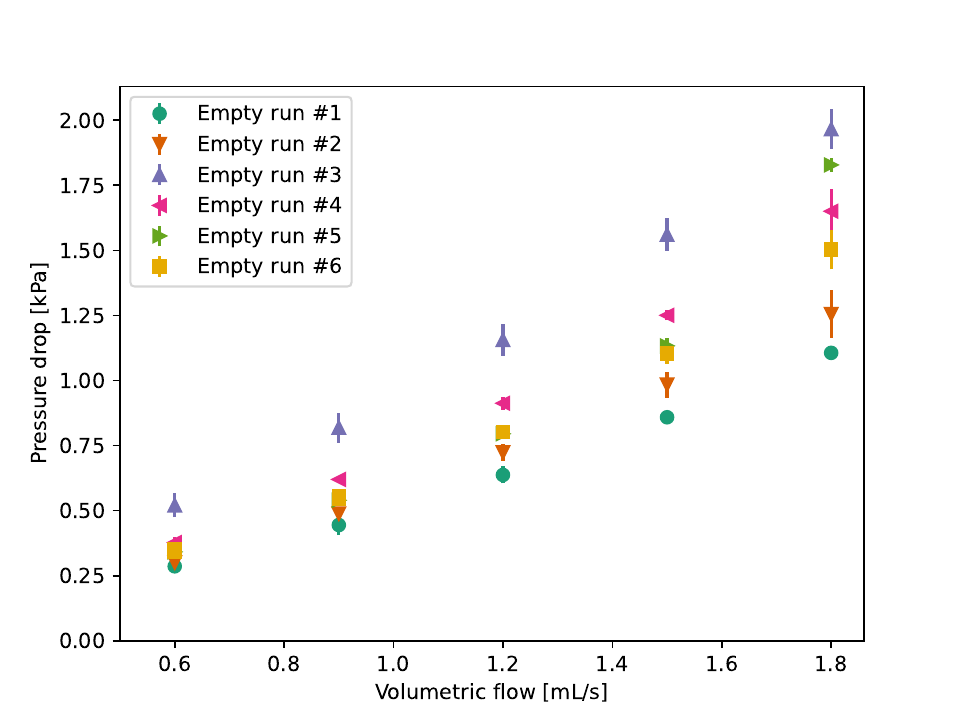}
        \caption{Pressure drop data for experiments where no sample is used. These pressure drops are to be subtracted from the experiments with samples to obtain the porous samples contributions. Each point is an average of three measures and confidence intervals are computed using a T-Student distribution, with a significance level $\alpha=0.05$.}
        \label{fig:blank_runs}
    \end{figure}

The data for the blank runs is shown in \textbf{Figure \ref{fig:blank_runs}}: at least three blank runs were done after each manipulation of the experimental setup. Each of the six triplets is presented as a confidence interval using a T-Student distribution, with a significance level $\alpha=0.05$. 
Each sequence of pressure measurements used for validation is associated to a blank run triplet which is subtracted from the associated measurements. Six blank runs are presented because experiments were executed in six sessions.
The variations are due to the sensor itself and  displacement of components of the experimental setup. To minimize the impacts of such variations, we calibrated the pressure sensor before each experiment session and ensured that component movement was minimal by securing the tubing to prevent changes in shape or positioning.

\section{Adaptive refinement of the mesh}
\label{sec:appendix_adaptive_3d}

\begin{figure}[htpb!]
        \centering
        \includegraphics[width=0.9\textwidth]{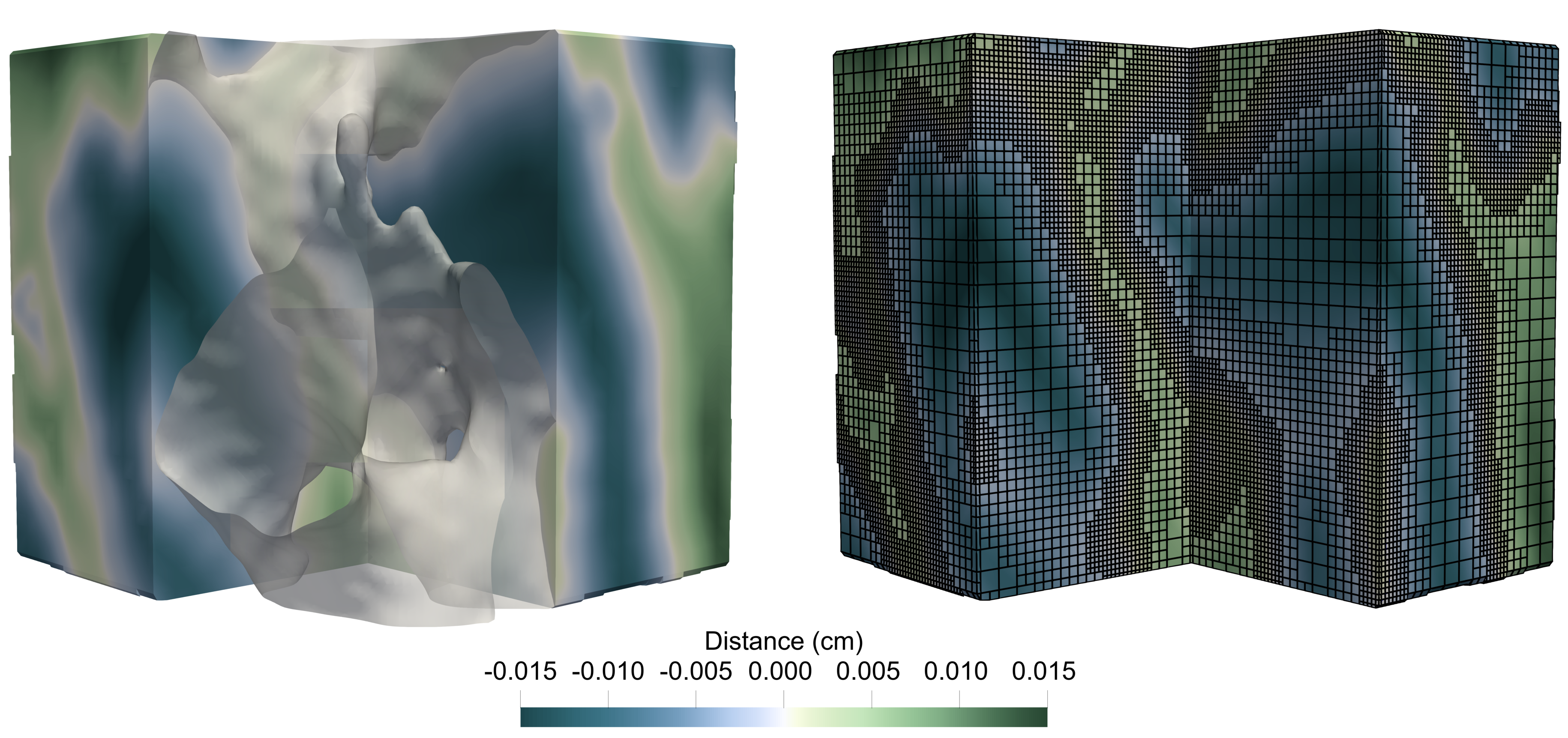}
        \caption{Subsection of a pore of sample \#2, with visible adaptively refined grid.}
        \label{fig:3d_adaptive_refinement}
    \end{figure}

\textbf{Figure \ref{fig:3d_adaptive_refinement}} shows part of a pore of sample \#2, with a subsection of its adaptively refined grid. 

\section{Validation with silicone porous media including results from all threshold values}
\label{sec:appendix_validation_threshold99}

\begin{figure}[htpb!]
        \centering
        \includegraphics[width=0.92\textwidth]{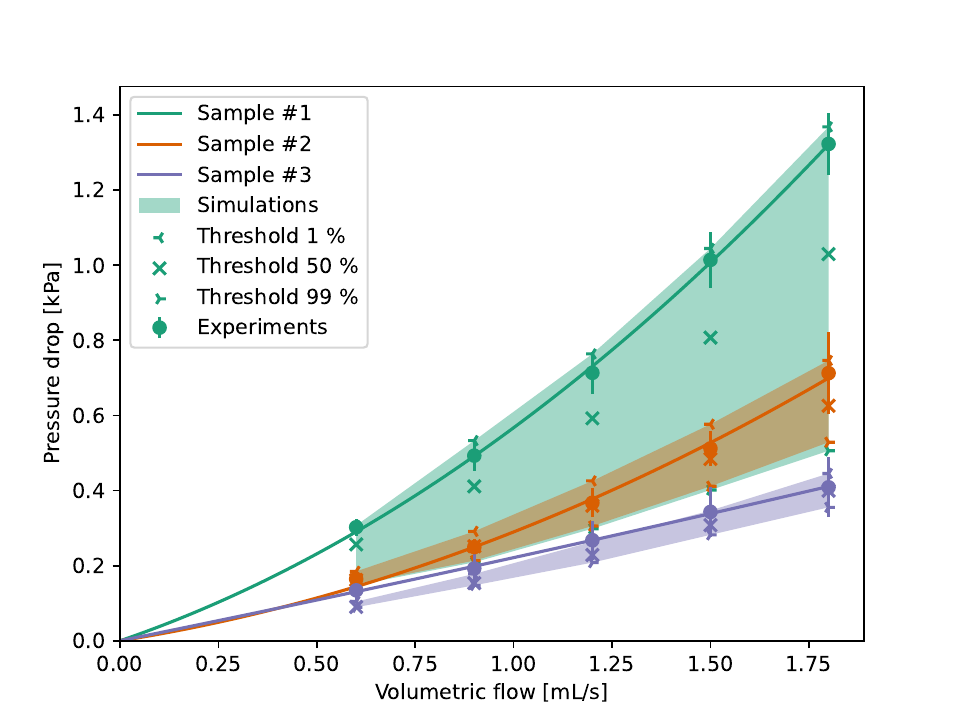}
        \caption{
        Experimental and simulated pressure drops of three silicone porous monoliths (average pore size: 300 $\mu$m). Darkened zones represent the values covered by thresholds between 1 \% and 99 \%. Experimental error bars are the T-Student's confidence interval using $\alpha = 0.05$ (n = 5 number of repetitions).}
        \label{fig:validation_with_99}
    \end{figure}

Simulated pressure drops through the digitized silicone monoliths are presented in \textbf{Figure \ref{fig:validation_with_99}}. The results include CFD simulation based on surface meshes generated using different thresholds $\{1,50,99\}$ \%, as well as previously shown experimental data. Sample \#1 is extremely sensitive to the threshold value, likely due to a combined effect of low mean pore size and standard deviation (see Table \ref{tab:samples_characterization}) and steadily low local porosity (see Figure \ref{fig:local_poro_simple}). The artificial occlusion or patency depending on the threshold parameter has stronger effects on this sample because of its smaller and more homogeneous pores.

\end{document}